%% file: main-sigconf.tex
\RequirePackage[hyphens]{url}

\documentclass[sigconf]{acmart}

\usepackage[keeplastbox]{flushend}					
\usepackage{multirow}
\usepackage[hyphens]{url}
\usepackage[para,online,flushleft]{threeparttable}
\usepackage{enumitem}

\newcommand{\nsf}[1]{\href{https://www.nsf.gov/awardsearch/showAward?AWD_ID=#1}{#1}}

\renewcommand\footnotetextcopyrightpermission[1]{}	
\usepackage{enumitem}
\usepackage{soul}
\usepackage{array}
\usepackage{tablefootnote}
\usepackage[normalem]{ulem}
\usepackage{makecell}
\usepackage{multirow}
\newcolumntype{C}[1]{>{\centering\let\newline\\\arraybackslash\hspace{1pt}}m{#1}}
\newcolumntype{L}[1]{>{\raggedright\let\newline\\\arraybackslash\hspace{1pt}}m{#1}}
\newcolumntype{R}[1]{>{\raggedleft\let\newline\\\arraybackslash\hspace{1pt}}m{#1}}

\AtBeginDocument{%
  \providecommand\BibTeX{{%
    \normalfont B\kern-0.5em{\scshape i\kern-0.25em b}\kern-0.8em\TeX}}}

\begin{document}
\let\hl\relax
%
\title{Survey of Transient Execution Attacks}

%
%
\author{Wenjie Xiong}
\affiliation{%
  \institution{Dept. of Electrical Engineering\\Yale University}
}
\email{wenjie.xiong@yale.edu}

\author{Jakub Szefer}
\affiliation{%
  \institution{Dept. of Electrical Engineering\\Yale University}
}
\email{jakub.szefer@yale.edu}
%
%
%
%
%


%

%
\begin{abstract}
\input{abstract}
\end{abstract}

%
%
%

%

%

%
\maketitle

\input{intro}

\input{components}

\input{transient_exe}

\input{covert_channel}

\input{transient_attacks}

\input{mitigations}

\input{conclusion}

\section*{Acknowledgements}
This work was supported in part by NSF grants \nsf{1651945} and \nsf{1813797},
and through SRC award number 2844.001.


%

%
\bibliographystyle{ACM-Reference-Format}
\bibliography{ref}

%

\end{document}

%% file: abstract.tex
Transient execution attacks, also called speculative execution attacks, 
\hl{have drawn much interest in the last few years as they can cause critical  data leakage.}
Since the first disclosure of transient execution attacks in January 2018, 
a number of new attack types or variants have been demonstrated  in different processors.
\hl{A transient execution attack consists of two main components: transient execution itself 
and a covert channel that is used to actually exfiltrate the information.}
\hl{Transient execution is caused by fundamental features of modern processors that boost performance and efficiency,
while covert channels are unintended channels that can be abused for information leakage, resulting from sharing of the micro-architecture components.}
Given the severity of the transient execution attacks,
they have motivated computer architects in both industry and academia to rethink the design of the processors and to propose hardware defenses.
\hl{To help understand 
the transient execution attacks, this paper summarizes the phases of the attacks and the security boundaries that are broken by the attacks.}
\hl{This paper further analyzes possible causes of transient execution 
and different types of covert channels. 
This paper in addition presents metrics for comparing different aspects of the transient execution attacks
(security boundaries that are broken, required control of the victim's execution, etc.) and uses them to
evaluate the feasibility of the different attacks -- both the existing attacks,
and potential new attacks suggested by our classification used in the paper.}
The paper finishes by discussing the different mitigations at the micro-architecture level that have so far been proposed.

%% file: intro.tex
\section{Introduction}
\label{sec:intro}

%
%

In the past decades, computer architects have been working hard to improve the performance of computing systems.
Different optimizations have been introduced in the various processor micro-architectures to improve the performance,
including pipelining, out-of-order execution, and branch prediction~\cite{hennessy2011computer}.
Some of the optimizations require aggressive speculation of the executed instructions.
For example, while waiting for a conditional branch to be resolved, branch prediction will predict whether the branch will be taken or not,
and the processor begins to execute code down the predicted control flow path before the outcome of the branch is known.
Such speculative execution of instructions causes the micro-architectural state of the processor to be modified.
The execution of the instructions down the incorrect speculated path is called the {\em transient execution} -- because
the instructions execute transiently and should ideally disappear with no side-effects if there was mis-speculation.
When a mis-speculation is detected, the architectural and micro-architectural side effects should be cleaned up -- but it is not done so today, leading to a number of recently publicized transient execution attacks~\cite{Kocher2018spectre,Lipp2018meltdown,canella2018systematic,van2018foreshadow,weisse2018foreshadow,ridl,fallout,Schwarz2019ZombieLoad} \hl{that leak data across different security boundaries in computing systems.}

%
%

Today's processor designs aim to ensure the execution of a program results in architectural states as if each instruction is
executed in the program order.
At the Instruction Set Architecture~(ISA) level, today's processors behave correctly.
Unfortunately, the complicated underlying micro-architectural states, due to different optimizations, are modified during the transient execution, and the various transient execution attacks have shown that data can be leaked from the micro-architectural states.
For example, timing channels can lead to information leaks that can reveal some of the micro-architectural states which are not visible
at the ISA level~\cite{szefer2018survey,liu2015last,he2017secure,zhang2014cross}.
\hl{Especially, the micro-architectural states of a processor 
are today not captured by the ISA specification, and there are micro-architectural vulnerabilities that cannot be found or analyzed by only examining the processor's ISA.}

%
%

Besides focusing on pure performance optimization, many processors are designed to share hardware units in order to reduce area and improve power efficiency.
For example,
hyper-threading allows different programs to execute concurrently on the same processor pipeline by sharing
the execution and other functional units among the hardware threads in the pipeline.
Also, because supply voltage does not scale with the size of the transistors~\cite{neamen2012semiconductor},
modern processors use multi-core designs.
In multi-core systems, caches, memory-related logic, and peripherals are shared among the different processor cores.
Sharing of the resources has led to numerous timing-based side and covert channels~\cite{szefer2018survey,liu2015last,he2017secure,zhang2014cross} -- the channels can occur independent of transient execution, or together with transient execution, which is the focus of this survey. 

%
%

\begin{figure*}[t]
\includegraphics[width=4.5in]{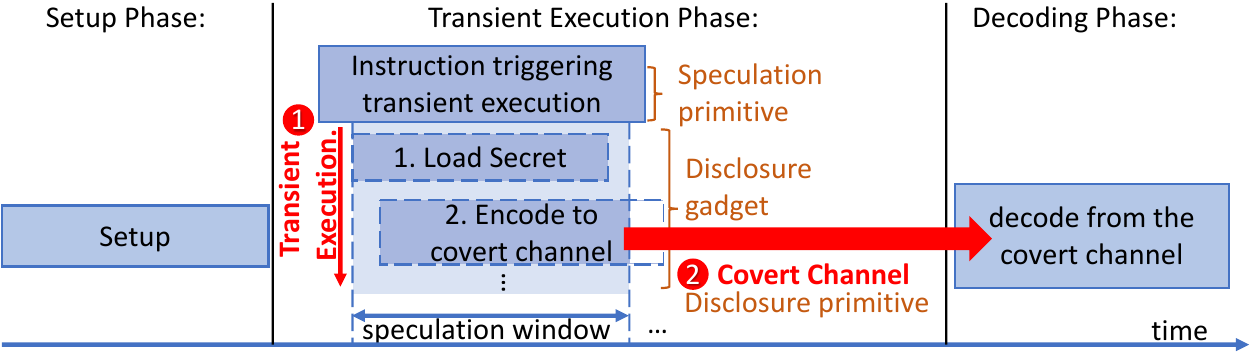}
\caption{\small Phases of transient execution attacks. }
 \label{fig:attack_phase}
\end{figure*}

Transient execution combined with covert channels results in {\em transient execution attacks} which can
compromise the confidentiality of the system. As shown in Figure~\ref{fig:attack_phase},
during such attacks, the secret or sensitive data is available during transient execution
-- this differentiates the transient execution attacks from conventional covert channel attacks where the data is assumed to be always available to the sender, not just during transient execution\footnote{There are also attacks using the timing difference in transient execution, e.g., \cite{evtyushkin2018branchscope,evtyushkin2015covert,evtyushkin2016jump,evtyushkin2018branchscope,Islam2018SPOILER}. These attacks are still conventional covert channel attacks, where the timing difference comes from the prediction units. Thus, these attacks are not in the scope of this paper, but are listed in Section~\ref{sec:related_attacks}.}. 
After the secret data is accessed during transient execution and encoded into a covert channel, the secret data
can be later extracted by the attacker from the covert channel.
 
A number of transient execution attack variants has been demonstrated, e.g.,
Spectre~\cite{Kocher2018spectre,maisuradze2018ret2spec,koruyeh2018spectre,chen2018sgxpectre,schwarz2018netspectre,v4,bhattacharyya2019smotherspectre,trippel2018meltdownprime}, Meltdown~\cite{Lipp2018meltdown,canella2018systematic,kiriansky2018speculative,v3a},
Foreshadow~\cite{van2018foreshadow,weisse2018foreshadow}, LazyFP~\cite{stecklina2018lazyfp},
Micro-architectural Data Sampling (MDS)~\cite{ridl,fallout,Schwarz2019ZombieLoad}, Load Value Injection (LVI)~\cite{van2020lvi}.
These attacks have been shown to allow data leaks across different security boundaries, e.g., system privilege level, SGX enclave, sandbox, etc.
\hl{The transient execution attacks have been assigned 9 Common Vulnerabilities and Exposures (CVE) IDs out of 14 CVE IDs that correspond to vulnerabilities about gaining information on Intel products in 2018, and 4 out of 9 in 2019, according to the CVE Details database}~\cite{cvedetails}. \hl{These attacks also affect other vendors, such as AMD or Arm, for example. }

In addition, these attacks have raised a lot of interest, and motivated computer architects to rethink the design of processors and propose a number of hardware defenses~\cite{yan2018invisispec,khasawneh2018safespec,kiriansky2018dawg,schwarz2019context,fustos2019spectreguard,barber2019isolating} -- this survey summarizes the attacks and the hardware defenses, while software-based defenses are summarized in existing work~\cite{canella2018systematic}.

\subsection{Outline and Contributions}

This paper provides a survey of existing {\em transient execution attacks} from Jan. 2018 to July 2020.
\hl{We start by providing background on the micro-architectural features that lead to the attacks.
We then define the transient execution attacks and summarize the phases and attack scenarios.
We analyze the types of transient execution and covert channels leveraged by the transient execution attacks to show the root causes of these attacks.}
In the end, we discuss the mitigation strategies for the transient execution and covert channels.
The contributions of this survey are the following:
\begin{itemize}
\item We summarize different attack scenarios and summarize the security boundaries that are broken by the attacks.
\item We provide a taxonomy of the existing transient execution attacks by analyzing the causes of transient execution that they leveraged,
and we propose metrics to compare the feasibility of the attacks.
\item We summarize and categorize the existing and potential timing-based covert channels in micro-architectures that can
be used with transient execution attacks, and also propose metrics to compare these covert channels.
\item We discuss the feasibility of the existing attacks based on the metrics we propose.
\item We compare the different mitigation strategies that have been so far designed at the micro-architectural level in various publications.
\end{itemize}


%% file: components.tex
\section{Transient Execution Attack Scenarios}
\label{sec:components}

We define transient execution attacks as attacks that access data during transient execution
and then leverage a covert channel to leak information.
The phases of these attacks are shown in Figure~\ref{fig:attack_phase}. 
\hl{
Although not indicated in the ``transient execution attacks" name, covert channels are an essential component of the transient execution attacks, because the micro-architectural states changed during transient execution are not visible at the architectural level, and are only accessible by using a covert channel to learn the state change (and thus the secret).} 
\hl{In this section, we summaries the attack scenarios, e.g., the attacker's goal, the location of the attacker, etc.}

\subsection{Attacker's Goal: Breaking Security Boundaries}
\label{sec:attacker_goal}

There are many security boundaries (between different privilege levels or security domains) in a typical processor, as shown in Figure~\ref{fig:security_boundary}.
The goal of the attacker of the transient execution attacks is to cross the security boundaries to obtain information related to the victim's protected data. 
In Figure~\ref{fig:security_boundary}, we categorize the possible privilege levels or security domains where the attack can originate and wherefrom it is trying to extract data as follows:
\begin{enumerate} 
\item \textbf{Across user-level applications:}
The attacker and the victim are two separate user applications, \hl{and the attacker process tries to learn the memory content of another process, e.g.,}~\cite{bhattacharyya2019smotherspectre}\hl{ demonstrates how an attacker process learns the private key when a victim OpenSSH server process is running in.}

\item \textbf{User-level program attacking the kernel:}
The attacker runs in the user level and wants to read the privileged data of the kernel, \hl{e.g., }\cite{Lipp2018meltdown}\hl{ demonstrates an attack that allows an unprivileged application to dump kernel memory.}

\item \textbf{Virtual machine attacking another virtual machine:}
The attacker and the victim resides in two different guest virtual machines, \hl{e.g.,} \cite{bhattacharyya2019smotherspectre}\hl{ shows it is possible for an attacker VM to learn the private key of OpenSSH server in the victim VM.} 

\item \textbf{Virtual machine attacking the hypervisor:}
\hl{The attacker is a guest OS and the victim is the host hypervisor, e.g.,}~\cite{Kocher2018spectre}\hl{ demonstrates an attack against KVM that leaks hypervisor's memory when the attacker has full control of the OS inside a VM.}

\item \textbf{Attacking the victim running inside an enclave:}
The victim runs inside a security domain protected by some hardware scheme, e.g., SGX enclaves~\cite{costan2016intel}, XOM~\cite{lie2000architectural}, Aegis~\cite{suh2014aegis}, Bastion~\cite{champagne2010scalable}, Sanctum~\cite{costan2016sanctum} or Keystone~\cite{lee2019keystone}, and the attacker code runs outside of it,
e.g.,~\cite{chen2018sgxpectre} \hl{demonstrates such an attack that retrieves secret from inside the SGX enclave.}

\begin{figure}[t]
\includegraphics[width=1.8in]{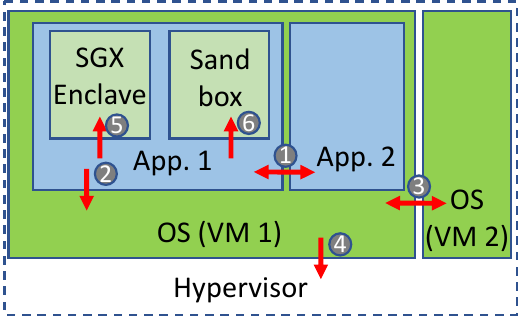}
\caption{\small Security boundaries in computer systems that are broken by transient execution attacks.}
\label{fig:security_boundary}
\end{figure}

\item \textbf{Across security domains protected by software:}
The victim runs inside the security domain protected by some software scheme, e.g., sandboxes in JavaScript, and the attacker code runs outside of it, as shown in~\cite{Kocher2018spectre}.

\end{enumerate}

\hl{All of the security boundaries listed above are broken by one or more of the existing transient execution attacks. The attacks have been shown to be able to retrieve coherent data, as well as  non-coherent data.  Details will be discussed in Section}~\ref{sec:consequence}, especially in Table~\ref{tbl:privilege_level}.

\subsubsection{Attacks Targeting Coherent and Non-Coherent Data}

\hl{We categorize all the data in the processor state into \textbf{coherent data} and \textbf{non-coherent data}. \textit{Coherent data} are those coherent with the rest of the system, e.g., data in caches are maintained by cache coherence protocol. Coherent data can be accessed by its address. \textit{Non-coherent data} are temporarily fetched into micro-architectural buffers or registers, are not synchronized with the rest of the system, and may not be cleaned up after use, e.g., data in the STL buffer. Thus, non-coherent data may be {stale}.
Non-coherent data that is left in the buffer can be of a different privilege level or security domain, so the attacker will break the security domain when accessing the non-coherent stale data.} Some attacks~\cite{v4,stecklina2018lazyfp} 
focus on attacking buffers to retrieve {such non-coherent data}, which in turn  breaks the security boundaries.

\subsection{Phases of the Attack}
As shown in Figure~\ref{fig:attack_phase}, we divide the transient execution attacks into three phases:

{\bf Setup Phase}: The processor executes a set of instructions that modify the micro-architectural
states such that it will later cause the transient execution of the desired code (called {\em disclosure gadget}) to occur in a manner predictable to the attacker.  An example is
performing indirect jumps to a specific address to ``train'' the branch predictor.  
The setup can be done by the attacker running some code or the attacker causing the victim to run in a predictable manner so that the micro-architectural state is set up as the attacker expects.

{\bf Transient Execution Phase}: The transient execution is actually triggered in this phase, and the desired disclosure gadget executes due to the prior training in the setup phase. \hl{The piece of code that accesses and transmits secret into the covert channel is called }{\em disclosure gadget}, following the terminology in~\cite{spectre_term}. The instructions belonging to the disclosure gadget are eventually squashed, and the architectural states of the transient instructions are rolled back, but as many of the attacks show, the micro-architectural changes caused by the disclosure gadget remain, so secret data can be later decoded from the covert channel.
This phase can be either executed by the victim or by the attacker.

{\bf Decoding Phase}: The attacker is able to recover the data via the covert channel by running the attacker's code or by triggering the victim's code and observing the behavior or result of the execution.

\hl{During an attack, the \textit{Setup Phase} and the \textit{Transient Execution Phase} cause the transient execution of the disclosure gadget to occur. Then, the \textit{Transient Execution Phase} and the \textit{Decoding Phase} leverage the covert channel to transmit data to the attacker. Thus, the Transient Execution Phase is critical for both accessing the secret and encoding it into a channel.}

\subsection{Transient Execution by the Victim vs. the Attacker}

\begin{figure*}[t]
\includegraphics[width=5.5in]{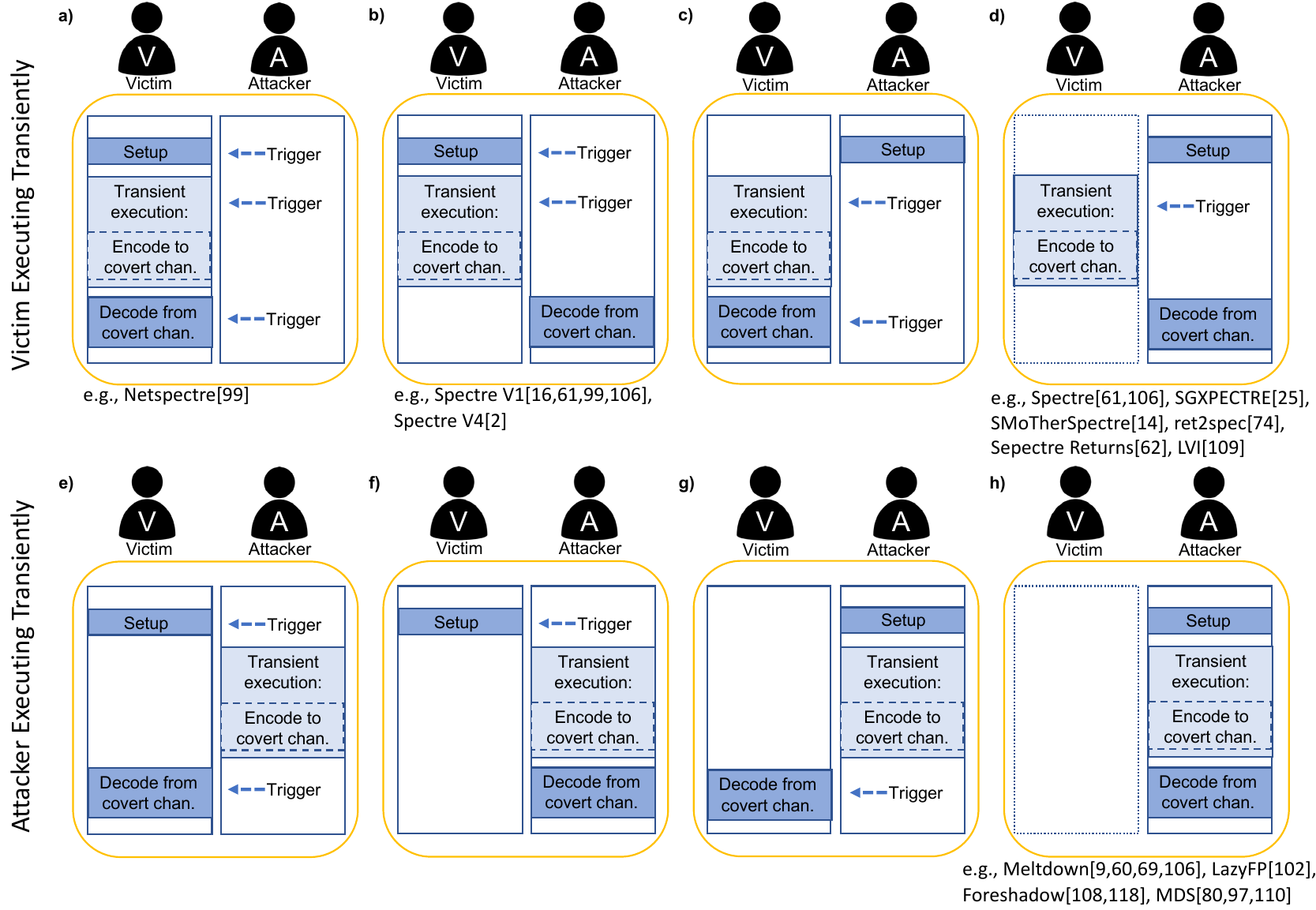}
\caption{\small  Possible scenarios of transient execution attacks:
a-d) the attacker triggers part of the victim code to execute transiently to leak secret through the covert channel, or
e-h) the attacker executes transiently to access data that she does not have permission to access and encodes it into the covert channel.
}
\label{fig:attacker_loc}
\end{figure*}

Each phase listed above can be performed by the attacker code or by the victim code, resulting in eight attack scenarios in Figure~\ref{fig:attacker_loc}. When a phase is performed by the victim, the attacker is assumed to have the ability to trigger the victim to execute the disclosure gadget.
We categorize the attacks based on who is executing transiently to encode the secret into the covert channel. 

\subsubsection{Victim is Executing Transiently.}


If the victim is the one who executes transiently, as shown in Figure~\ref{fig:attacker_loc} (a--d), the victim is triggered to execute a disclosure gadget that encode some secret into the covert channel during transient execution, and the attacker obtains the secret by decoding the data from the covert channel. In this scenario, the attacker is assumed to be able to control or trigger the execution of the disclosure gadget in the victim's codebase. The attacker can do this by calling some victim functions with certain parameters. For example, in SGXpectre~\cite{chen2018sgxpectre}, the attacker can launch the target enclave~program.

Different from the conventional side and covert channels, here, the encoding phase is executed transiently, and thus, the attack cannot be detected by simply analyzing the software semantics of the victim code. This attack vector leverages the difference between the expected semantics of software execution and the actual execution in hardware and is a fundamental problem in current computer architectures.

There are two options for preparing for the transient execution (i.e., setup phase). First, if the hardware component that causes transient execution, e.g., the prediction unit, is shared between the attacker and the victim, then the attacker's execution can manipulate the prediction unit to trigger the disclosure gadget in the victim code to execute transiently, as shown in Figure~\ref{fig:attacker_loc} (c,d). The second option is that  the attacker triggers a setup gadget in the victim codebase to set up the transient execution, as shown in Figure~\ref{fig:attacker_loc} (a,b).
For the first option, the attacker is required to share the prediction unit with the victim and to prepare some code to set up the hardware to lure the victim into desired transient execution.
For the second option, the attacker is required to understand the victim's code and be able to trigger the setup gadget to execute with a controlled input, e.g., by calling a function of the victim code.

Decoding data from the covert channel can be done by the attacker code, as shown in Figure~\ref{fig:attacker_loc} (b,d), or by the victim code, as shown in Figure~\ref{fig:attacker_loc} (a,c). For the second case, the attacker may directly query a decoding gadget in the victim code and leverage the results of the decoding gadget the infer information through the covert channel, or the attacker may trigger the execution of the decoding gadget and measure the time or other side effect of the execution.

\subsubsection{Attacker is Executing Transiently.}
As shown in Figure~\ref{fig:attacker_loc} (e--h), the attacker can directly obtain the secret in transient execution.
The attacker will then encode the data into a covert channel and decode it to obtain the secret in the architectural state, such as in her memory. 
The attacker can also launch different software threads for the setup or the decoding phases. 
The attacker's code shown in Figure~\ref{fig:attacker_loc} (e--h) might be in different threads, even on different cores.

During the attack, the attacker directly obtains the secret during transient execution, and thus, the attacker should be able to have a pointer to the location of the victim data. 
There might be only the attacker code running, or the attacker and the victim running in parallel.
When there is only the attacker code running, the victim's protected data should be addressable to the attacker or the data is in some register in the hardware, i.e., the attacker should have a way to point to the data.
In Meltdown~\cite{Lipp2018meltdown}, the attacker code first loads protected data by its virtual address to register and then transfers the data through a covert channel.
When the attacker and the victim are running concurrently, the attacker should be able to partially control the victim's execution or synchronize with the victim execution.
For example, in Micro-architectural Data Sampling (MDS) attacks~\cite{ridl,fallout,Schwarz2019ZombieLoad}, the attacker needs to synchronize with the victim execution to extract useful information from the non-coherent data of the victim in the buffers.

In micro-architectural implementations, transient execution allows the attacker to access more data than it is allowed in the architecture (ISA) level. Thus, this type of attack is implementation-dependent and does not work on all the CPUs, e.g., Meltdown~\cite{Lipp2018meltdown}, Foreshadow~\cite{van2018foreshadow,weisse2018foreshadow}, MDS~\cite{ridl,fallout,Schwarz2019ZombieLoad}, are reported to work on Intel processors.

Similar to the case when the victim is executing transiently, the setup phases and decoding phases can also be done by the victim, resulting in four attack scenarios in Figure~\ref{fig:attacker_loc} (e--h). However, in the current known attacks, the attacker always sets up, triggers the transient execution, and decodes from the channel, which is more practical.

\subsubsection{\hl{Feasibility of the Attack Scenarios}}
\label{sec:feasibility_scenario}

\begin{table*}[t]
\centering
\caption{\small \hl{Required Control of Victim Execution in Different Attack Scenarios.}}
\begin{threeparttable}
\small
\begin{tabular}{ |C{5em} || C{5em} | C{5em} |C{5em} || C{5em} | C{5.5em} | C{5em}  | }\hline
\textbf{Scenario in Figure~\ref{fig:attacker_loc}} & \textbf{Setup Phase}  & \textbf{Transient Execution Phase} & \textbf{Decoding Phase}  & \textbf{Number of Victim Gadgets to~be Triggered*} & \textbf{Sharing Required during Transient Execution**} & \textbf{Sharing Required for Covert Channel**}\\
\hline
a& Victim & Victim & Victim & 2-3 & No&No\\
b& Victim & Victim & Attacker & 1-2 & No&Yes\\
\hline
c& Attacker & Victim & Victim & 2 &Yes&No\\
d& Attacker & Victim & Attacker &1  &Yes&Yes\\
\hline
\hline
e& Victim & Attacker & Victim & 2 &Yes&Yes\\
f& Victim & Attacker & Attacker & 1 &Yes&No\\
\hline
g& {Attacker} & Attacker & Victim & 1 &No&Yes\\
h& {Attacker} & Attacker & Attacker & 0 & No& No\\
\hline
\end{tabular}
* The number shows the number of different code gadgets in the victim's codebase to be triggered by the attacker. 
We assume the decoding gadget is different from the disclosure gadget. The setup gadget may or may not be the same code as the disclosure gadget, 
so the two gadgets can be counted as either $1$ (same) or $2$ (different) gadgets,
giving a range of gadgets required, as show in the fifth column of the table.

** Here, we refer to sharing of hardware between the attacker and the victim. In addition, the attacker (or the victim) could also have multiple software threads running and sharing hardware between the threads. We assume colocation between the each party's threads is possible, and do not list that here.
\begin{tablenotes}
\end{tablenotes}
\end{threeparttable}
\label{tbl:pred_ctrl_victim}
\end{table*}

\hl{The required number of gadgets in the victim codebase to be triggered and required sharing in different transient execution scenarios is summarized in Table}~\ref{tbl:pred_ctrl_victim}. In addition, Figure~\ref{fig:attacker_loc} \hl{shows the attack scenarios demonstrated in different publications. 
In a practical attack, it is desired to have most phases to be executed by the attacker's code and less required sharing of hardware.}

\hl{In most of the existing attacks,
the attacker completes setup and decoding steps, as shown in Figure}~\ref{fig:attacker_loc} \hl{(d,h), because they use less gadgets in the victim codebase and are more practical for the attacker. Attack scenarios (a,b) in Figure}~\ref{fig:attacker_loc} \hl{are also demonstrated that have less requirement of shared hardware.}
In Spectre V1, since the victim disclosure gadget can be reused as the setup gadget for training the predictor,  triggering victim to run the setup phase does not require additional effort for the attacker, and thus, Figure~\ref{fig:attacker_loc}~(b) is also practical.
The attacker can also use the victim's code to complete both setup and decoding steps, as shown in Figure~\ref{fig:attacker_loc} (a). In this case, the attacker can launch the attack remotely~\cite{schwarz2018netspectre}. 


Scenarios (c) and (e--g) in Figure~\ref{fig:attacker_loc}\hl{ require more gadgets in the victim code and are not demonstrated in the publications so far. However, if the attacker has the ability to trigger the victim to execute certain gadgets (as required by some of the attacks already), those scenarios are still feasible and should be considered when designing mitigations.}

%% file: transient_exe.tex
\section{Transient Execution}
\label{sec:transient_exe}

\hl{Transient execution is the phenomenon where code is executed speculatively, and it is not
known if the instructions will be committed or squashed until the retirement of the instruction or a pipeline squash event.
Upon the squash, 
not all the micro-architectural side effects are cleaned up properly, causing the possible 
transient execution attacks. 
Hence, all causes of pipeline squash are also causes of transient execution and need to be understood to know what cause transient execution attacks to occur.
In this section, we first discuss all possible causes of transient execution, 
then we propose a set of the metrics to evaluate feasibility of the transient 
execution attacks.}

\begin{table*}[t]
\centering
\caption{\small Data Leaked by the Transient Execution Attacks.}
\begin{threeparttable}
\centering
\small
\begin{tabular}{ l | l | l | l| l|c c c c c c |c  }
&\multicolumn{3}{c|}{\multirow{5}{4.5em}{\bf Causes of Transient Execution}}  &\multicolumn{1}{c|}{\multirow{5}{4em}{\bf Example Attacks}} &  \multicolumn{6}{c|}{\bf Coh. Data**} &  \multirow{2}{2em}{{\bf Non-coh. Data**}}\\
 &\multicolumn{3}{l|}{} && \rotatebox{90}{hypervisor} &  \rotatebox{90}{across VM} & \rotatebox{90}{kernel data} &  \rotatebox{90}{across user app.} &  \rotatebox{90}{SGX} &  \rotatebox{90}{sandbox}   &  \rotatebox{90}{}  \\
\hline
\multirow{10}{4.5em}{{Victim Executes {Transiently}}} 
& \multirow{6}{2em}{{\rotatebox{90}{Prediction}}}&\multirow{3}{2em}{{Ctrl Flow}}&PHT&
Spectre V1~\cite{Kocher2018spectre,schwarz2018netspectre,trippel2018meltdownprime,swapgs}& $\boxtimes$ & $\boxtimes$ &$\boxtimes$ & $\boxtimes$ & $\boxtimes$ &$\boxtimes$ & $\Box$\\
&&&BTB&Spectre V2~\cite{Kocher2018spectre,chen2018sgxpectre,bhattacharyya2019smotherspectre}& $\boxtimes$ & $\boxtimes$ &$\boxtimes$ & $\boxtimes$ & $\boxtimes$ &$\boxtimes$ & $\Box$\\
&&&RSB&Spectre V5~\cite{maisuradze2018ret2spec,koruyeh2018spectre}& $\boxtimes$ & $\boxtimes$ &$\boxtimes$ & $\boxtimes$ & $\boxtimes$ &$\boxtimes$ & $\Box$\\
\cline{3-12}
&&\multirow{2}{1.5em}{Addr.}& \multirow{1}{1em}{STL}&Spectre V4, LVI~\cite{v4,van2020lvi} &$\boxtimes$ & $\boxtimes$ &$\boxtimes$ & $\boxtimes$ & $\boxtimes$ &$\boxtimes$ & $\boxtimes$\\
&&&LFB&LVI~\cite{van2020lvi}&$\boxtimes$ & $\boxtimes$ &$\boxtimes$ & $\boxtimes$ & $\boxtimes$ &$\boxtimes$& $\boxtimes$\\
\cline{3-12}
&&Value& \multicolumn{8}{l}{no commercial implementation} \\
\cline{2-12}
&\multicolumn{2}{l|}{{Exception}}& *&LVI~\cite{van2020lvi}&$\boxtimes$ & $\boxtimes$ &$\boxtimes$ & $\boxtimes$ & $\boxtimes$ &$\boxtimes$ & $\boxtimes$\\
\cline{2-12}
&\multicolumn{3}{l|}{{Interrupts}}& \multicolumn{8}{l}{no known attack}\\
\cline{2-12}
&\multicolumn{3}{l|}{{Load-to-load reordering}}& \multicolumn{8}{l}{no known attack}\\
\hline
\multirow{12}{4.5em}{{Attacker Executes Transiently}} &  
 \multirow{5}{2em}{{\rotatebox{90}{Prediction}}}
&\multirow{1}{2em}{{Ctrl Flow}} 
&*& \multicolumn{8}{l}{no known attack}\\ 
&&&\\
\cline{3-12}
&& \multirow{2}{2em}{Addr.}
& STL&Fallout~\cite{fallout} &$\Box$ &$\Box$ &$\Box$ &$\Box$ &$\Box$ &$\Box$ & $\boxtimes$\\
&&& LFB&RIDL, ZombieLoad~\cite{ridl,Schwarz2019ZombieLoad} &$\Box$ &$\Box$ &$\Box$ &$\Box$ &$\Box$ &$\Box$ & $\boxtimes$ \\
\cline{3-12}
&&Value& \multicolumn{8}{l}{no commercial implementation} \\
\cline{2-12}
& \multicolumn{2}{l|}{\multirow{5}{4em}{Exception}}& PF-US &Meltdown (V3)~\cite{Lipp2018meltdown,trippel2018meltdownprime} &$\Box$ &$\Box$ & $\boxtimes$ & $\Box$&$\Box$ &$\Box$ & $\Box$\\
&\multicolumn{2}{l|}{}& PF-P& Foreshadow (L1TF)~\cite{van2018foreshadow,weisse2018foreshadow} &$\boxtimes$ &$\boxtimes$ &$\boxtimes$ &$\boxtimes$ &$\boxtimes$ &$\Box$ &$\Box$ \\
&\multicolumn{2}{l|}{}& PF-RW&V1.2~\cite{kiriansky2018speculative} &$\Box$ &$\Box$ &$\Box$ &$\Box$ &$\Box$ &$\boxtimes$ &$\Box$ \\
&\multicolumn{2}{l|}{}& NM&LazyFP ~\cite{stecklina2018lazyfp} &$\Box$ &$\Box$ &$\Box$ &$\Box$ &$\Box$ &$\Box$ &$\boxtimes$ \\
&\multicolumn{2}{l|}{}& GP&V3a~\cite{v3a}&$\Box$&$\Box$ &$\boxtimes$  &$\Box$ &$\Box$ &$\Box$ &$\Box$ \\
\cline{2-12}
&\multicolumn{3}{l|}{{Interrupts}}& \multicolumn{8}{l}{no known attack}\\
\cline{2-12}
&\multicolumn{3}{l|}{{Load-to-load reordering}}& \multicolumn{8}{l}{no known attack}\\
\hline
\end{tabular}
\begin{tablenotes}
$\boxtimes$ indicates that the attack can leak the protected data; $\Box$ indicates that the attack cannot leak the data.\\
* indicates all hardware components that cause the corresponding transient execution, we combine them in the same row because the data leaked in the attacks are the same.

**{\em Coh. Data} is short for coherent data, {\em Non-coh. Data} is short for non-coherent data.
\end{tablenotes}
\end{threeparttable}
\label{tbl:privilege_level}
\label{tbl:spec_primitive}
\end{table*}

\subsection{Causes of Transient Execution}

The following is an exhaustive list of possible causes of transient execution (i.e., causes of  pipeline squashing).

\textbf{Mis-prediction:} The first possible cause of transient execution is mis-prediction. Modern computer architectures make predictions to make full use of the pipeline to gain performance. When the prediction is correct, the execution continues and the results of the predicted execution will be used. In this way, predictions boost performance by executing instructions earlier. If the prediction is wrong, the code (transiently) executed down the incorrect (mis-predicted path) will be squashed. 
There are three types of predictions: control flow prediction, address speculation, and value prediction.

\begin{enumerate}[itemindent=24pt,leftmargin=0pt,listparindent=\parindent]
\item \textbf{Control Flow Prediction:} Control flow prediction predicts the execution path that a program will follow.
Branch prediction unit (BPU) stores the history of past branch directions and targets and leverages the locality in the program control flow to make predictions for future branches. 
BPU predicts whether the branch is to be taken or not (i.e., branch direction) by using pattern history table (PHT), and what is the target address (i.e., branch or indirect jump target) by using branch target buffer (BTB) or return stack buffer (RSB).  The implementation details of PHT, BTB, and RSB in Intel processors will be discussed in Section~\ref{sec:Ctrl_flow_Intel}.

\item \textbf{Address Speculation:}
Address speculation is a prediction on the address when the physical address is not fully available yet, e.g., whether two addresses are the same. It is used to improve performance in the memory system, e.g., store-to-load (STL) forwarding in the load-store queue, line-fill buffer (LFB) in the cache. The implementation details of STL and LFB in Intel processors will be discussed in Section~\ref{sec:Addr_spec_Intel}.

\item \textbf{Value Prediction:}
To further improve the performance, while the pipeline is waiting for the data to be loaded from memory hierarchy on a cache miss, value prediction units have been designed to predict the data value and to continue the execution based on the prediction.  
While this is not known to be implemented in commercial architectures, value prediction had been proposed in the literature~\cite{lipasti1996value,lipasti1996exceeding}. 

\end{enumerate}

\textbf{Exceptions:} 
The second possible cause for transient execution to occur are exceptions.
\hl{If an instruction causes an exception, the handling of the exception is sometimes delayed until the instructing is retired,  allowing code to (transiently) execute until the exception is handled. There are a number of causes of exceptions, such as a wrong permission bit (e.g., present bit, reserved bit) in Page Table Entry (PTE), etc.} A list of all the exception types or permission bit violations is summarized in \cite{canella2018systematic}. \hl{In addition, Xiao et al. developed a software framework to automatically explore the vulnerabilities on a variety of Intel and AMD processors}~\cite{xiao2019speechminer}.

Sometimes the exceptions are suppressed due to another fault, e.g., nested exceptions. For example, when using transactional memory (Intel TSX~\cite{TSX}), if a problem occurs during the transaction, all the architectural states in the transaction will be rolled back by a transaction abort, suppressing the exception that occurred in the middle of the transaction~\cite{ridl,Schwarz2019ZombieLoad}. Another way is to put the instruction that would cause exception in a mis-predicted branch.
In this survey, even if the exception is suppressed later, we categorize the attack to be due to exceptions.

\textbf{Interrupts:} The third possible cause for transient execution is (external) interrupts. If a peripheral device or a different core causes an interrupt, \hl{the processor stops executing the current program, saves the states, and transfers control to interrupt handler. In one common implementation, when stoping execution, the oldest instruction in the ROB will finish execution, and all the rest of the instructions in the ROB will be squashed, the instructions that were executed after the oldest instruction (but end up being squashed) are executed transiently.}
After the interrupt is handled, the current program may continue the execution, i.e., the instructions that are squashed will be fetched into the pipeline again.

\textbf{Load-to-Load Reordering (Multi-Core):} 
The fourth possible cause for transient execution is load-to-load reordering. Current x86 architectures use the total store order~(TSO) memory model~\cite{sewell2010x86}.  \hl{In TSO, all observable load and store reordering are not allowed except store to load reordering where a load bypasses an older store of a different address.} To prevent a load to load reordering, if a load has executed but not yet retired and the core receives a cache invalidation for the line read by the load, the pipeline will be squashed. Transient execution occurs between the instruction issue and when the load-to-load reordering is detected.


\subsection{Causes of Transient Execution in Known Attacks}
Not all transient execution can be leveraged in an attack,
and Table~\ref{tbl:spec_primitive} shows the causes of transient execution in existing attacks.
Mis-prediction is leveraged in Spectre-type attacks, e.g.,~\cite{Kocher2018spectre}. Address speculation is leveraged in MDS attacks~\cite{ridl,fallout,Schwarz2019ZombieLoad} and LVI~\cite{van2020lvi}. Exceptions of loads or stores are leveraged in Meltdown attacks~\cite{Lipp2018meltdown}, Foreshadow attacks~\cite{van2018foreshadow,weisse2018foreshadow}, and LVI~\cite{van2020lvi}, etc.
Other types of exceptions, interrupts, and load-to-load reordering are not considered to be exploitable. Because the instructions that get squashed due to exceptions, interrupts and load-to-load, are legal to be resumed later on, and no  extra data is accessible to the attacker during the transient execution.

\begin{figure*}[t]
\includegraphics[width=4.5in]{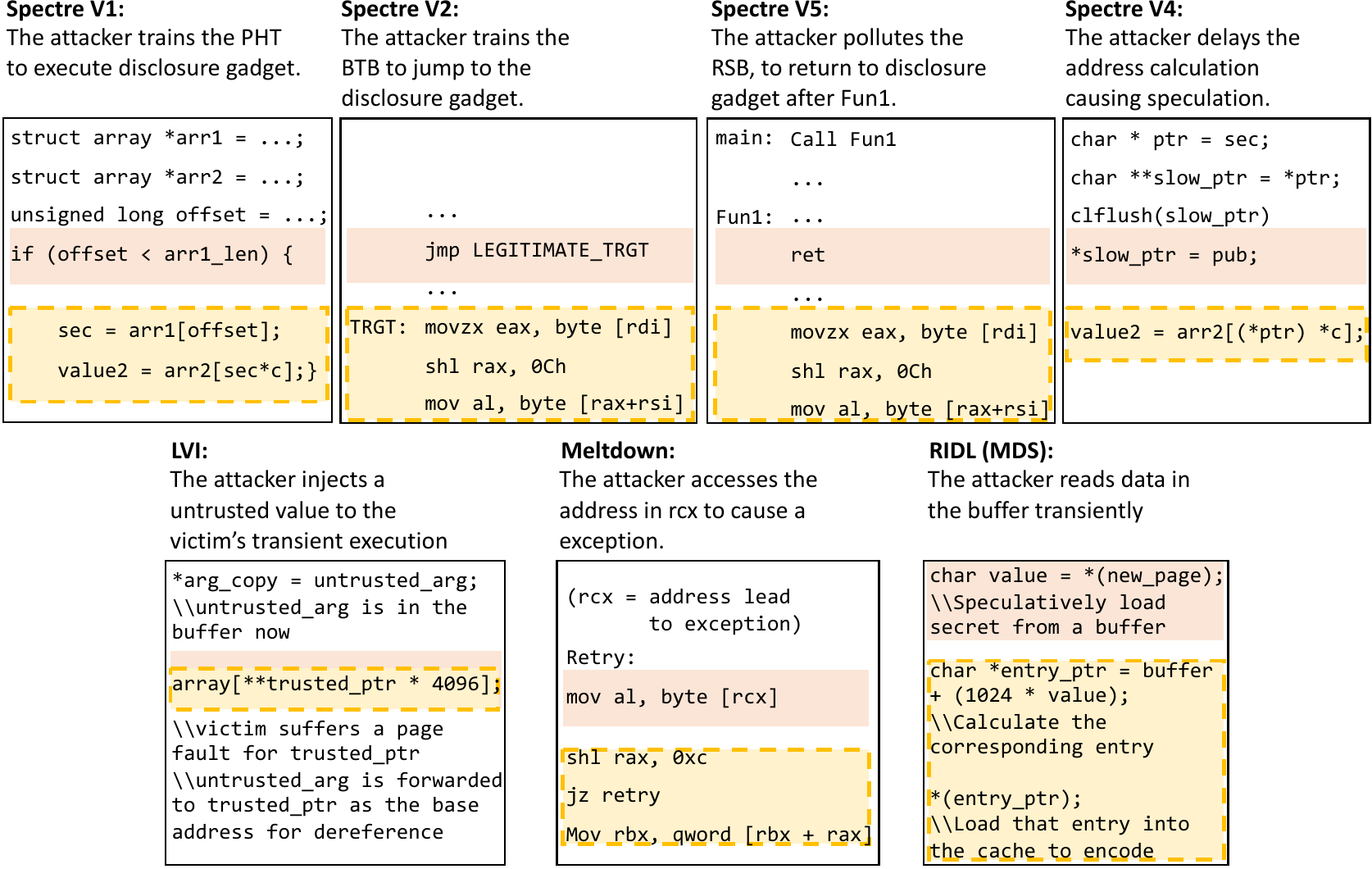}
\caption{\small Example code of transient execution attacks. Code highlighted in orange triggers transient execution. Code highlighted in yellow with dashed frame is the disclosure gadget.}
 \label{fig:spectre_sample}
\end{figure*}

The sample codes of different variants are shown in Figure~\ref{fig:spectre_sample}. The victim code should allow a potential mis-prediction or exception to happen. In Spectre V1~\cite{Kocher2018spectre}, to leverage PHT, a conditional branch should  exist in the victim code followed by the gadget. Similarly, in Spectre V2~\cite{Kocher2018spectre} and V5~\cite{maisuradze2018ret2spec,koruyeh2018spectre}, the victim code should have an indirect jump (or a return from a function) that uses BTB (or RSB) for prediction of the execution path. In Spectre V4~\cite{v4}, to use STL, the victim code should have a store following a load having potential address speculation.
\hl{In LVI}~\cite{van2020lvi}\hl{, a load that triggers a page fault (accessing} {\tt trusted\_ptr}\hl{) will forward non-coherent data in the store buffer which is injected by a malicious store (}{\tt *arg\_copy = untrusted\_ptr}\hl{), and then, the secret data addressed by the injected value (}{\tt **untrusted\_ptr}\hl{) is leaked.
In Meltdown}~\cite{Lipp2018meltdown}\hl{, the attacker code should make  an illegal load to cause an exception. 
In MDS attack}~\cite{ridl,Schwarz2019ZombieLoad,fallout}\hl{, a faulty load (}{\tt value=*(new\_page)}\hl{) will forward non-coherent data in the buffer.}

\subsection{Metrics for Causes of Transient Execution}
\label{sec:transient_metric}
If the attacker wants to launch a transient execution attack, the attacker should be able to cause transient execution of the disclosure gadget in a controlled manner.
We propose the following metrics to evaluate the different causes of transient execution:
\begin{itemize}[itemindent=15pt,leftmargin=0pt,listparindent=\parindent]
\item \textbf{Security Boundaries that are Broken}:
This metric indicates the security boundaries that are broken during the transient execution attacks -- this will be discussed in Section~\ref{sec:consequence}.
\item \textbf{Required Control of  the Victim's Execution}:
This metric evaluates whether the attacker needs to control the execution of victim code  -- details will be discussed in Section~\ref{sec:mistrain_setup}.
\item \textbf{Required Level of Sharing}:
This metric evaluates how close the attacker should co-locate with the victim and whether the attacker should share memory space with the victim to trigger the transient execution in a controlled manner  -- details will be discussed in Section~\ref{sec:mistrain}.
\item \textbf{Speculative Window Size}:
This metric indicates how many instructions can be executed transiently -- the speculation window size will be discussed in more detail in Section~\ref{sec:spec_window}.
\end{itemize}

\subsection{Security Boundaries that are Broken}
\label{sec:consequence}

As discussed in Section~\ref{sec:attacker_goal}\hl{, the attacker's goal is to access the coherent or non-coherent data across the security boundaries in the system.} 
Table~\ref{tbl:privilege_level} \hl{lists the type of data and the security boundaries across which the data can be leaked in the known transient execution attacks, assuming all the instructions in the disclosure gadget can execute transiently and the covert channel can transmit information to the attacker.}

\hl{If the victim is executing transiently, the disclosure gadget can read any coherent data that the victim could access architecturally, 
even if the semantics of the victim code do not intend it to access the data}~\cite{Kocher2018spectre}. 
\hl{Hence, in these attacks, the attacker can break the isolation between the victim and the attacker and learn data in the victim's domain.}
For example, SWAPGS instruction is a privileged instruction that usually executed after switching from user-mode to kernel-mode. If SWAPGS is executed transiently in the kernel-mode in the incorrect path, kernel data can be leaked~\cite{swapgs}.
\hl{When the victim is executing transiently, the attacker can also learn the non-coherent data (for example, stale data) and also data that depends on non-coherent data (e.g., data in an address that is depended on non-coherent data).
For example, in Spectre V4}~\cite{v4}, \hl{stale data that contains the address of the secret data in the store buffer is forwarded to the younger instructions transiently, and the disclosure gadget accesses and transmits the secret data to the attacker. 
As another example, in LVI attack}~\cite{van2020lvi}\hl{, the attacker injects malicious value through buffers, such as STL or LFB, causing a victim's transient execution that depends on a value controlled by the attacker and potentially leaks the value in address controlled by the attacker.} 

\hl{If the attacker is executing transiently, transient execution allows the attacker to access illegal data directly.} 
As shown in Table~\ref{tbl:privilege_level}\hl{, the security boundaries that are broken depend on the causes of transient execution.} 
In some processor implementations, even if a load causes an exception due to permission violation, the coherent data might still be propagated to the following instructions and  learned by the attacker.
For example, in Meltdown~\cite{Lipp2018meltdown}, privileged data is accessible transiently to an unprivileged user even if the privileged bit in the page table is set.
In L1 terminal fault (L1TF)~\cite{weisse2018foreshadow}, secret data in the L1 cache is accessible transiently even if the present bit in the page table is not set.
In Table~\ref{tbl:privilege_level}, the attacks leveraging exceptions are categorized by the cause of the exception, e.g., page fault (PF), and the related permission bit. 
\hl{Non-coherent data present in the micro-architecture buffers (e.g., Line Fill Buffers (LFB) or store buffer (STB)) can sometimes be accessed by the attacker in transient execution}~\cite{ridl,Schwarz2019ZombieLoad,fallout}. In addition, in CROSSTALK~\cite{ragabcrosstalk}, \hl{a hardware buffer called staging buffer is discovered. The staging buffer is for some type of off-core reads, e.g., {\tt RDRAND} instruction that requesting DRNG (Digital Random Number Generator), {\tt CPUID} instruction that read from MachineSpecific Registers (MSRs).
The staging buffer is shared across cores, and thus,  the CROSSTALK paper demonstrated a cross-core attack where the victim fetch some data from RNG, and the attacker then learn the random number in the stage buffer during transient execution.}

\subsection{Required Control of the Victim's Execution}
\label{sec:mistrain_setup}


For the attacks leveraging mis-prediction, (mis-)training is a essential setup step to steer the control flow to execute the desired disclosure gadget.
The (mis-)training can be part of victim code, which is triggered by the attacker, as shown in Figure~\ref{fig:attacker_loc} (b) and Table~\ref{tbl:pred_ctrl_victim}. In the example of Spectre V1, the attacker can first provide inputs to train the branch predictor (i.e., PHT) to execute the gadget branch, because in this way the training code will always share the branch predictor with the attack code. In this case, the attacker should be able to control the execution of victim code.
The (mis-)training code can also be a part of the attacker's code and run in parallel with the victim code, as shown in Figure~\ref{fig:attacker_loc} (d), e.g., in Spectre V2. Then, it is required that the attacker's training thread and the victim's thread should be co-located to share the same prediction unit (e.g., BTB). Further, to share the same entry of the prediction unit, if the prediction unit is indexed by physical address, the attacker and the victim should also share the same memory space to share the entry, which will be discussed in the next subsection.

For the attacks that leverage exceptions, the instructions that follow the exception will be executed transiently, and thus, no mis-training is required, but the attacker needs to make sure the disclosure gadget is located in the code such that it is executed after the exception-causing instruction.

\subsection{Required Sharing during Transient Execution}
\label{sec:mistrain}

\hl{As shown in Table}~\ref{tbl:pred_ctrl_victim}\hl{, in some scenarios, the setup code and the disclosure gadget are run by different parties, e.g., Figure~}\ref{fig:attacker_loc}\hl{ (c-f), or in attacker's different software threads, e.g., Figure}~\ref{fig:attacker_loc}\hl{ (g-h). These cases require that the setup code shares the same prediction unit (entry) with the disclosure gadget. 
One common attack scenario is that the attacker mis-trains the prediction unit to lure the execution of the disclosure gadget of the victim, e.g., Figure}~\ref{fig:attacker_loc}\hl{ (d).
Hardware sharing can be as follows:} 

\label{sec:colocation_attacker}
%
\begin{itemize}
\item \textbf{Same thread:}  The attacker and the victim  (if both of them executing) or the attacker's software threads (if only the attacker is executing) are running on the same logical core (hardware thread) in a time-sliced setting, and there might be  context switches in between.
\item \textbf{Same core, different thread:}   The attacker and the victim  (if both of them executing) or the attacker's threads (if only the attacker is executing) are running on different logical cores (hardware threads) through simultaneous multithreading (SMT) on the same physical core.
\item \textbf{Same chip, different core:}  The attacker and the victim  (if both of them executing) or the attacker's threads (if only the attacker is executing) are on different CPU cores, but are sharing LLC, memory bus, and other peripheral devices.
\item \textbf{Same motherboard, different chip:}  The attacker and the victim  (if both of them executing) or the attacker's threads (if only the attacker is executing) share memory bus and peripheral devices.

\end{itemize}

\label{sec:addr_space_attacker}
\hl{Some prediction units have multiple entries indexed by address, and in that case, the attacker needs to share the same entry of the prediction unit with the victim during the setup. To share the same entry, the attacker needs to control the address to map to the same predictor entry as the victim.}
The address space can be one of the following:
\begin{itemize}
\item \textbf{In the same address space:} In this case, the attacker and the victim have the same virtual to physical address mapping.
\item \textbf{In different address spaces with shared memory:} In this case, the attacker and the victim have different virtual to physical address mappings. But some of the attacker's pages and the victim's pages map to the same physical pages. This can be achieved by sharing dynamic libraries (e.g., {\tt libc}).
\item \textbf{In different address spaces without shared memory:} The attacker and the victim have different virtual to physical address mapping.  Further, their physical addresses do not overlap.
\end{itemize}

\begin{table*}[t]
\centering
\caption{\small Level of Sharing and (Mis-)training the Prediction Unit  on Intel Processors.}
\begin{threeparttable}
\small
\begin{tabular}{  l l | l l l l  }
&  \parbox{0.7in}{\bf Prediction Unit} &\parbox{0.5in}{\rotatebox{30}{\bf same thread}} &  \parbox{0.5in}{\rotatebox{30}{\bf same core, different thread}} &  \parbox{0.6in}{\rotatebox{30}{\bf same chip, different core}} &  \parbox{0.8in}{\rotatebox{30}{\bf same motherboard}}    \\
\hline
\multirow{3}{0.4in}{\rotatebox{0}{Ctrl Flow}}  & PHT~\cite{evtyushkin2018branchscope,kiriansky2018speculative} & f(virtual addr) &f(virtual addr) &-- & -- \\
&BTB~\cite{Kocher2018spectre,evtyushkin2016jump} & {f(virtual addr)} & {f(virtual addr)}\tnote{a}   & -- & -- \\
&RSB~\cite{maisuradze2018ret2spec} & not by address\tnote{b}& -- &-- &-- \\
\hline
\multirow{3}{0.4in}{\rotatebox{0}{Addr.}}  &STL~\cite{Islam2018SPOILER,fallout} & f(physical addr) \tnote{c}  & -- & --& -- \\
&LFB~\cite{ridl,Schwarz2019ZombieLoad} & not  by address  & not  by address & --& -- \\
&Other\tnote{d} &  &  & & \\
\hline
Value & no commercial impl.  &  &  & & \\
\hline
\end{tabular}
\begin{tablenotes}
\footnotesize
``--" indicates the prediction unit is not possible to be trained under the corresponding sharing setting;
Otherwise, the prediction unit can be trained and
``f(virtual addr)" indicates the prediction unit is indexed by a function of the virtual address,
``f(physical addr)" indicates the prediction unit is indexed by a function of the  physical address,
 and ``not  by address" indicates the prediction unit is not indexed by addresses.\\
\item[a] Conflicting results are presented in different publications~\cite{evtyushkin2016jump,Kocher2018spectre}. \\
\item[b] Most OSes overwrite RSBs on context switches.\\
\item[c] STL is possible after context switch, but not on SGX enclave exit.\\
\item[d] In~\cite{Schwarz2019ZombieLoad}, it is indicated that there could be other structures which forward data speculatively.
\end{tablenotes}
\end{threeparttable}
\label{tbl:pred_share}
\end{table*}

\hl{In the following, we discuss the level of sharing required to trigger transient execution of disclosure gadget for an attack leveraging mis-prediction. 
In particular, the scenario depends on the implementation, and thus, we discuss each of the prediction units in Intel Processors in detail.}

\subsubsection{Control Flow Prediction:} 
\label{sec:Ctrl_flow_Intel}
To predict the branch direction, modern branch predictors use a hybrid mechanism~\cite{mcfarling1993combining,evers1996using,michaud1997trading,sprangle1997agree,jimenez2001dynamic}. One major component of the branch predictor is the pattern history table (PHT). Typically, a PHT entry is indexed based on some bits of the branch address, so a branch at a certain virtual address will always use the same entry in the PHT. In each entry of the PHT, a saturating counter stores the history of the prior branch results, which in turn is used to make future predictions.

To predict the branch targets, a branch target buffer (BTB) stores the previous target address of branches and jumps. Further, a return instruction is a special indirect branch that always jumps to the top of the stack. The BTB does not give a good prediction rate on return instructions, and thus, return stack buffer (RSB) has been introduced in commercial processors. 
The RSB stores $N$ most recent return addresses.

In Intel processors, the PHT and BTB\footnote{In~\cite{evtyushkin2016jump}, the authors did not observe BTB collision between logical cores.  However, it is demonstrated that the attacker can mis-train the indirect jump of a victim when they are two hyper-threads sharing the same physical core in~\cite{Kocher2018spectre}. Thus, we think BTB is shared across hyper-threads in some of the processors. }
 are shared for all the processes running on the same physical core (same or different logical core in SMT).
The RSB is dedicated to each logical core in the case of hyper-threading~\cite{maisuradze2018ret2spec}.
Table~\ref{tbl:pred_share} shows whether the prediction unit can be trained when the training code and the victim are running in parallel in different settings.
The results are implementation-dependent and Table~\ref{tbl:pred_share} shows the result from Intel processors.

The prediction units sometimes have many entries, and the attacker and the victim should use the same entry for mis-training.
The attacker and the victim will use the same entry only if they are using the same index. When the prediction unit is indexed by virtual address, the attacker can train the prediction unit from another address space using the same virtual address as the victim code. If only part of the virtual address is used as the index, which is shown as {\tt f(virtual addr)} in Table~\ref{tbl:pred_share}, the attacker can even train with an aliased virtual address, which maps to the same entry of the prediction unit as the victim address.
The RSB is not indexed by the address, rather it overflows when many nested calls are made,
and this creates conflicts when there are more than $N$ nested calls, and will cause mis-prediction.

\begin{figure*}[t]
\includegraphics[width=4in]{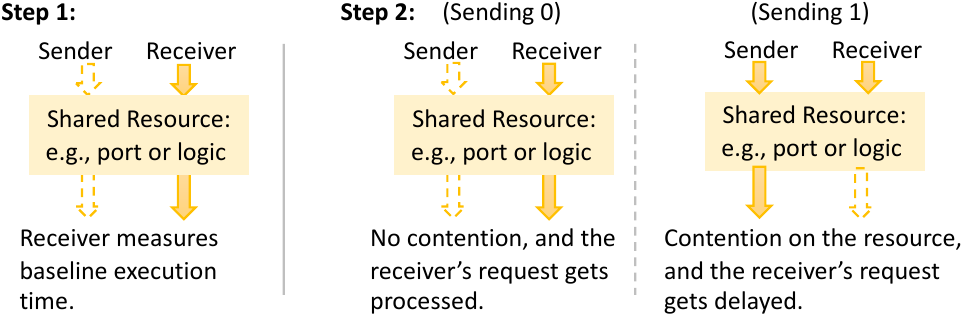}
\caption{\small Steps for the sender and the receiver to transfer information through volatile covert channels. The yellow box shows the shared resource. The solid (dashed) arrow shows the shared resource is (is not) requested or used by the corresponding party.}
 \label{fig:volatile_channel}
\end{figure*}

\subsubsection{Address Speculation:} 
\label{sec:Addr_spec_Intel}
One of the uses of address speculation is in the memory disambiguation to resolve read-after-write hazards, which are the data dependencies between instructions in out-of-order execution.
In Intel processors, there are two known uses of address speculation. First, loads are assumed not to conflict with earlier stores with unknown addresses, and speculatively store-to-load (STL) forwarding will not happen. When the address of a store is later resolved, the addresses of younger loads will be checked. And if store-to-load forwarding should have happened and data dependence has been violated, the loads will be flushed, and the new data is reloaded from the store, as shown in the attacks~\cite{v4,schwarz2019store}. Second, for performance, when the address of a load {\em partially} matches the address of a preceding store, the store buffer will forward the data of the store to the load speculatively,  even though the full addresses of the two may not match~\cite{fallout}. In the end, if there is mis-prediction, the load will be marked as faulty, flushed, and reloaded again.

Another use of address speculation is in conjunction with the line-fill buffer (LFB), which is the buffer storing cache-lines to be filled to the L1 cache. LFB may forward data speculatively without knowledge of the target address~\cite{ridl,Schwarz2019ZombieLoad}.
Address speculation may also be used in other hardware structures in Intel processors, as indicated in ~\cite{Schwarz2019ZombieLoad}.

To trigger address speculation, the availability of the address should be delayed to force the hardware to predict the address.
One way is to make the address calculation depends on some uncached data, as in Spectre V4~\cite{v4}.
Another way is to use a newly mapped page, so that the physical address is available only after OS handles the page-in event, as in~\cite{ridl}. In an extreme case, the speculation can even be caused by a NULL pointer or an invalid address, and then the error is suppressed in the attacker code, as in attack~\cite{Schwarz2019ZombieLoad}.
In STL, the entries are indexed by a function of physical addresses. In this case, the training code needs to share memory space with the victim to achieve an attack.

\subsubsection{Value Prediction:}  There is no commercial processor that implement value prediction yet. 
Thus, there are no known exploits that abuse value prediction. However, similar to control flow prediction, if the predictor is based on states that are shared between different threads and not cleaned up during context switch, the prediction can be hijacked by the attacker.

\subsection{Speculative Window Size}
\label{sec:spec_window}
To let an attack happen, there should be a large enough speculative window for the disclosure gadget to finish executing transiently, as shown in  Figure~\ref{fig:attack_phase}. The speculative window size is the window from the time the transient execution starts (instruction fetch) to the time the pipeline is~squashed.
In attacks leveraging predictions, the speculative window depends on the time the prediction is resolved. In a conditional branch, the time  depends on the time to resolve the branch condition; in indirect jump, this depends on the time to obtain the target address; and in address speculation, this depends on the time to get the virtual and then the physical address.
\hl{In}~\cite{mambretti2019speculator}\hl{, a tool called \textit{Speculator} is proposed to reverse engineer the micro-architecture using hardware performance counters. The results of the \textit{Speculator} show the speculative window of branches that depend on uncached data is about 150 cycles on Intel Broadwell, about 300 cycles on Intel Skylake, and about 300 cycles on AMD Zen, and the  speculative window of STL is about 55 cycles on Intel Broadwell.}
In attacks leveraging exceptions, the speculative window depends on the implementation of~exceptions.
To make the speculative window large enough for the disclosure gadget, the attacker can delay the obtaining of the result of the branch condition or the addresses by leveraging uncached loads from main memory, chains of dependent instructions, etc.

%% file: covert_channel.tex
\section{Covert Channels}
\label{sec:covert_channel}

Transient execution enables the attacker to access the secret data transiently,
and
a covert channel\footnote{The channel is considered a covert channel, not a side channel~\cite{Kocher2018spectre}, because the attacker has 
control over the disclosure gadget, which encodes the secret. }~\cite{szefer2018survey} is required
for the attacker to eventually obtain the secret data in architectural states.
There is a distinction between {\em conventional channels} where the encoding happens in software execution path,
and {\em transient execution channels} where the encoding phase is executed transiently.
Here, we focus on covert channels that can be used in transient attacks -- these can also be used as conventional covert channels.

There are two parties in a covert channel: the sender and the receiver. 
In the covert channels, the sender execution will change some micro-architectural state and the receiver will observe the change to extract information, e.g., by observe the execution time.
 
 \begin{table*}[t]
\setlength{\tabcolsep}{3pt}
\centering
\caption{\small Known Covert Channels in Micro-architecture.}
\centering
\begin{threeparttable}
\small
\begin{tabular}{  l l | l l l l  | c C{0.8in} }
 & \multirow{8}{1.3in}{\bf Covert Channel Type} &\multicolumn{4}{C{0.6in}|}{\bf Level of Sharing}   & \multirow{8}{0.6in}{\bf Bandwidth} & \multirow{8}{0.95in}{\bf Required Time~Resolution of the Receiver (CPU cycles)} \\
 &  &\parbox{0.1in}{\rotatebox{90}{same thread}} &  \parbox{0.1in}{\rotatebox{90}{same core, different thread}}  &  \parbox{0.1in}{\rotatebox{90}{same chip, different core}} &  \parbox{0.1in}{\rotatebox{90}{same motherboard}}  & &  \\
\hline
\multirow{4}{0.6in}{{Volatile Covert Channels}} & Execution Ports~\cite{wang2006covert,bhattacharyya2019smotherspectre,aldayaport} &$\boxtimes$ & $\boxtimes$ &$\Box$& $\Box$& not given & 50 vs. 80\\
& FP division unit~\cite{fustos2020spectrerewind} &$\boxtimes$ & $\boxtimes$ &$\Box$& $\Box$ & $\sim$70kB/s & 314 vs. 342\\
& L1 Cache Ports~\cite{yarom2017cachebleed,moghimi2018memjam} &$\boxtimes$ & $\boxtimes$ &$\Box$& $\Box$ & not given & 36 vs. 48\\
& Memory Bus~\cite{wu2014whispers} &$\boxtimes$ & $\boxtimes$ & $\boxtimes$& $\boxtimes$  & $\sim$700 B/s & 2500 vs. 8000\\

\hline
\multirow{10}{0.6in}{{Persistent Covert Channels}} 
& AVX2 unit~\cite{schwarz2018netspectre} &$\boxtimes$ & $\boxtimes$ &$\Box$& $\Box$ & $>$0.02B/s & 200 vs. 550\\
& PHT~\cite{evtyushkin2018branchscope} &$\boxtimes$ & $\boxtimes$ &$\Box$& $\Box$ & not given & 65 vs. 90\\
& BTB~\cite{evtyushkin2016jump,weisse2019nda} &$\boxtimes$ & $\boxtimes$ &$\Box$& $\Box$ & not given & 56 vs. 65\\
& STL~\cite{Islam2018SPOILER} &$\boxtimes$ & $\Box$ &$\Box$& $\Box$ & not given & 30 vs. 300\\
&TLB~\cite{gras2018translation,hund2013practical,schwarz2019store} &$\boxtimes$ & $\boxtimes$ &$\Box$& $\Box$ &  $\sim$5kB/s per set & 105 vs. 130\tnote{a}\\
&L1, L2 (tag, LRU)~\cite{xu2011exploration,kiriansky2018dawg,Xiong2019Leaking} &$\boxtimes$ & $\boxtimes$ &$\Box$& $\Box$ & $\sim$1MB/s per cache entry & 5 vs. 15\tnote{b} \\
&LLC (tag, LRU)~\cite{liu2015last,briongos2019reload} &$\Box$ & $\Box$  &$\boxtimes$ & $\Box$ &  $\sim$0.7MB/s per set & 500 vs. 800\\
&Cache Coherence~\cite{yao2018coherence,trippel2018meltdownprime}  &$\Box$ & $\Box$  &$\boxtimes$ & $\boxtimes$ & $\sim$1MB/s per cache entry & 100 vs. 250\tnote{c}\\
&Cache Directory~\cite{yan2019attack} &$\Box$ & $\Box$  &$\boxtimes$ & $\Box$ & $\sim$0.2MB/s per slice & 40 vs. 400\\
&DRAM row buffer~\cite{pessl2016drama}&$\Box$ & $\Box$ &$\boxtimes$ & $\boxtimes$ &  $\sim$2MB/s per bank& 300 vs. 350\\
\hline
\end{tabular}
\begin{tablenotes}
$\boxtimes$ indicates that the attack is possible to leak the protected data; $\Box$ indicates that the attack cannot leak the data.

\item[a] Depending on the level of TLB used, the required time resolution varies. The biggest one is shown.
\item[b] Shows the time resolution for covert channel use L1 cache.
\item[c] Depending on the setup, the required time resolution varies. The biggest one is shown.
\end{tablenotes}
\end{threeparttable}
\label{tbl:channel_share}
\end{table*}

\subsection{Assumptions about Covert Channels}
\label{sec:receiver_obv}
This survey focuses on covert channels that do not require physical presence and which only require attacker's software (or software under the attacker's control) to be executing on the same system as the victim.
Thus, we do not consider physical channels, such as power~\cite{genkin2015get}, EM field~\cite{matyunin2016covert}, acoustic signals~\cite{backes2010acoustic,genkin2014rsa}, etc.
There are certain physical channels that can be accessed from software and not require physical presence, such as temperature~\cite{xiong2019spying}. 
However, thermal conduction is slow and the bandwidth is limited.

Any sharing of hardware resources between users could lead to a covert channel between a sender and a receiver~\cite{wang2006covert}.
The receiver can observe the status of the hardware with some metadata from the covert channel, such as the execution time, values of hardware performance counters (HPC), system behavior, etc.

The most commonly used observation by the receiver of the covert channels is the timing of execution.
In today's processors, components are designed to achieve a better performance, and thus, the execution time contains information about whether certain hardware unit is available during execution (e.g., port), whether the micro-architectural states are optimal for the code (e.g., cache hits or misses), etc.
To observe the hardware states via timing, a timer is needed.  In x86, {\em rdtscp} instruction can be used to read a high-resolution time stamp counter of the CPU, and thus, can be used to measure the latency of a chosen piece of code. When the {\em rdtscp} is not available, a counting thread can be used as a timer~\cite{schwarz2017fantastic}.

The receiver can also gain information from hardware performance counters (HPCs). HPCs have information about branch prediction, cache, TLB, etc, 
and have been used in covert channel attacks~\cite{evtyushkin2018branchscope}.
However, HPCs must be configured in kernel mode~\cite{das2019sok}, and thus, are not suitable for unprivileged attackers.

The receiver can further observe the state of the hardware by the system behaviors.
In Prime+Abort attack~\cite{disselkoen2017prime}, for example, TSX can be exploited to allow an attacker to receive an abort (call-back) if
the victim process accessed a critical address. 

In other cases, several covert channels are used in series. Here, for transient execution attacks, we only consider channels where the receiver can decode data architecturally.
For example, in the Fetch+Bounce covert channel~\cite{schwarz2019store}, first, the secret is encoded into the TLB states, which affect the STL forwarding, and then a  cache Flush+Reload covert channel is used to observe the STL forwarding results.
The first channel can only be observed by instructions in transient execution and the states will be removed when the instruction retires.
We only consider the second covert channel to be critical for transient execution attack because 
it allows the attacker to observe the secret architecturally.


\subsection{Types of Covert Channels}

We categorize the covert channels into \textbf{volatile channels} and \textbf{persistent channels}.
In volatile channels, the sender and the receiver share the resource on the fly, no states are changed, e.g., sharing a port or some logic concurrently.  The sender and the receiver have contention when communicating using this type of channel. 
In persistent channels, the sender changes the micro-architectural states, and the receiver can observe the state changes later, e.g., change of cache state. Although the states may be changed later, we call them persistent channels to differentiate from the volatile channels. 
The persistent covert channels will be discussed in the next subsection.

\subsection{Volatile Covert Channels}
\label{sec:volatile_channel}

In a \textit{volatile covert channel}, there is contention for hardware between the sender and the receiver on the fly, and thus, the two should run concurrently, for example, as two hyper-threads in SMT processors, or running concurrently on two different cores. 
Another scenario is that the sender and the receiver are two part of code in the same software thread that their instructions are scheduled to execute concurrently due to OoO~\cite{fustos2020spectrerewind}. 
As shown in Figure~\ref{fig:volatile_channel}, the receiver first measures the baseline execution time when the sender is not using the shared resource. Then, the sender causes contention on the shared resource or not depending on the message to be sent, while the receiver continues to measure the execution time. If the execution time increases, the receiver knows the sender is using the shared resource at the moment.

Execution units, ports, and buses are shared between the hyper-threads running concurrently on the same physical core, and can be used for covert channels~\cite{bhattacharyya2019smotherspectre,aldayaport}.
\hl{There is also a covert channel leveraging the contention in the floating point division} unit~\cite{fustos2020spectrerewind}.
L1 cache ports are also shared among hyper-threads.
In Intel processors, L1 cache is divided into banks, and each cache bank can only handle a single (or a limit number of) requests at a time. CacheBleed~\cite{yarom2017cachebleed} leverages the contention L1 cache bank to build a covert channel. Later, Intel resolved the cache bank conflicts issue with the Haswell generation.
However, MemJam~\cite{moghimi2018memjam} attack demonstrates that there is still a false dependency of memory read-after-write requests when the addresses are of the same L1 cache set and offset for newer generations of Intel processors. This false dependency can be used for a covert channel.
As shown in Table~\ref{tbl:channel_share}, the covert channel in execution ports and L1 cache ports can lead to covert channels \hl{within the same thread when the sender and the receiver code are executed in parallel due to OoO} and between hyper-threads in SMT setting.

Memory bus serves memory requests to all the cores using the main memory.
In~\cite{wu2014whispers}, it is shown that the memory bus can act as a high-bandwidth covert channel medium, and covert channel attacks on various virtualized x86 systems are demonstrated.

\subsection{Persistent Covert Channels}
\label{sec:persistent_channel}

In a \textit{persistent channel}, the sender and the receiver share the same micro-architectural states, e.g., registers, caches, etc.
Different from volatile covert channels, the state will be memorized in the system for a while. And the sender and the receiver do not have to execute concurrently.
Depending on whether the state can only be used by one party or can be directly accessed by different parties in the system, we further divide the persistent channels into occupancy-based and encode-based, as shown in Figure~\ref{fig:persistent_channel}.


\subsubsection{Occupancy-based Persistent Covert Channels}
To leverage occupancy-based covert channel, the user needs to occupy the states (e.g., registers, cache, or some entries) or data to affect the execution.

\begin{itemize}[itemindent=15pt,leftmargin=0pt,listparindent=\parindent]
\item \textit{Eviction-based Persistent Channels:}
In this channel, the sender and the receiver will compete and evict the other party to occupy some states to store their data or metadata to (de-)accelerate their execution. One example of the eviction-based channel is the Prime+Probe attack~\cite{osvik2006cache,percival2005cache,guanciale2016cache,yan2019attack,yan2019cache}. The receiver first occupies a cache set (i.e., primes).  Then, the sender may use the state for her data or not, depending on the message to be sent. And in the end, the receiver reads (i.e., probes) her data that were used to occupy the cache set in the first step to see whether those data are still in the cache by measuring the timing, as shown in the first row of Figure~\ref{fig:persistent_channel}. Other examples of the eviction-based channel are cache Evict+Time attack~\cite{osvik2006cache,bernstein2005cache}, the covert channel in DRAM row buffer~\cite{pessl2016drama}.

\begin{figure*}[t]
\includegraphics[width=4.3in]{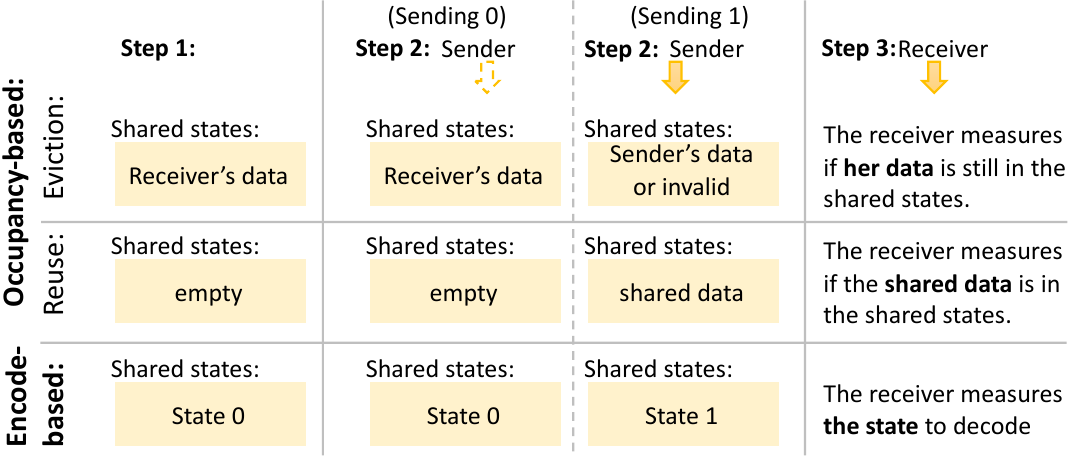}
\caption{\small Steps for the sender and the receiver to transfer information through different types of persistent covert channels.}
 \label{fig:persistent_channel}
\end{figure*}

Another possible contention is that the sender needs to use the same piece of data (e.g., need exclusive access to the data for write), and thus, the receiver's copy of data can be invalidated.
Some state is used for tracking the relationship of data in different components, which can cause the data in one component to be invalidated.
For example, cache coherency policy can invalidate a cache line in a remote cache, and thus, it results in a covert channel between threads on different cores on the same processor chip~\cite{yao2018coherence,trippel2018meltdownprime}.
Cache directory keeps the tags and cache coherence state of cache lines in the lower levels of cache in a non-inclusive cache hierarchy and can cause eviction of a cache line in the lower cache level (a remote cache relative to the sender) to build a covert channel~\cite{yan2019attack}.

\item \textit{Reuse-based Persistent Channels:}
In this channel, the sender and the receiver will share some data or metadata, and if the data is stored in the shared state, it could (de-)accelerate both of their execution. The cache Flush+Reload attack~\cite{yarom2014flush,gruss2015cache} transfers information by {\em reusing} the same data in the cache. The receiver first cleans the cache state. Then, the sender loads the shared data or not. And in the end, the receiver measures the execution time of loading the shared data, as in Figure~\ref{fig:persistent_channel}. If the sender loads the shared data in the second step, the receiver will observer faster timing compared to the case when the sender does not load the shared data.
There are other reuse-based attacks, such as Cache Collision attack~\cite{bonneau2006cache} and the cache Flush+Flush attack~\cite{gruss2016flush}.

\hl{Prediction units can also be leveraged for such covert channels due to a longer latency for mis-prediction. 
For example, PHT}~\cite{evtyushkin2018branchscope,evtyushkin2015covert,yu2019speculative}, BTB~\cite{evtyushkin2016jump,weisse2019nda}, and STL~\cite{Islam2018SPOILER} have been demonstrated to be usable for constructing covert channels. For example, when sharing BTB, the sender and the receiver use the same indirect jump source, ensuring the same BTB entry is used. If the receiver has the same destination address as the sender, the BTB will make a correct prediction resulting in a faster jump.

\end{itemize}

\subsubsection{Encode-based Persistent Covert Channels}
In encode-based persistent covert channels,  the sender and the receiver can both directly change and probe the shared state. One example of such a channel is the AVX channel~\cite{schwarz2018netspectre}. There are two AVX2 unit states: power-off and power-on.  To save power, the CPU can power down the upper half of the AVX2 unit by default. In step 2, if the sender then uses the AVX2 unit, it will be power-on the unit for at least 1 ms. In step 3, the receiver can measure whether the AVX2 unit is power-on by measuring the time of using AVXs unit. In this way, the sender encodes the message into the state of the AVX2 unit, as shown in Figure~\ref{fig:persistent_channel}.
Other examples are the covert channels leveraging cache LRU states~\cite{kiriansky2018dawg,briongos2019reload,Xiong2019Leaking}.


\subsection{Metrics for Covert Channels}
\label{sec:CC_metric}

We propose the following metrics to compare different covert channels:
\begin{itemize}[itemindent=15pt,leftmargin=0pt,listparindent=\parindent]
\item \textbf{Level of Sharing}:
This metric indicates how the sender and the receiver should co-locate.
As shown in Table~\ref{tbl:channel_share}, some of the covert channels only exists when the sender and the receiver share the same physical core. Other attacks exist when the sender and the receiver share the same chip or even the same motherboard.
\item \textbf{Bandwidth}:
This metric measures how fast the channel is. The faster the channel, the faster the attacker can transfer the secret.  Table~\ref{tbl:channel_share} compared the bandwidth of different covert channels. Usually, the bandwidth is measured in a real system considering the noise from activities by other software and the operating system.
\item \textbf{Time Resolution of the Receiver}:
As shown in Figures~\ref{fig:volatile_channel} and~\ref{fig:persistent_channel}, the receiver needs to measure and differentiate different states. For a timing channel, the time resolution of the receiver's clock decides whether the receiver can observe the difference between the sender sending 0 or 1. The last column of Table~\ref{tbl:channel_share} shows the timing difference between states. Some channels, such as cache L1, require a very high-resolution clock to differentiate 5 cycles from 15 cycles, while the LLC covert channel only needs to differentiate 500 cycles from 800 cycles, and the receiver only needs a coarse-grained clock.
\item \textbf{Retention Time}:
This metric measures how long the channel can keep the secret.
In some of the covert channels (volatile channels in Section~\ref{sec:volatile_channel}), no state is changed, e.g., the channel leveraging port contention~\cite{aldayaport}. The retention time of such channels is zero, and the receiver must measure the channel concurrently when the sender is sending information.
Other covert channels (persistent channels in Section~\ref{sec:persistent_channel}) leverage state change in micro-architecture, the retention time depends on how long the state will stay, for example, AVX2 unit will be powered off after about 1ms. If the receiver does not measure the state in time, she will obtain no information. For other states, such as register, cache, etc., the retention time depends on the usage of the unit and when the unit will be used by another user.
\end{itemize}

\subsection{Comparison of Covert Channels}
\label{sec:comp_CC}
Table~\ref{tbl:channel_share}\hl{ lists different covert channels in micro architecture.
The existence of covert channel depends on whether the unit is shared in that setting.}
For example, AVX2 units, TLB, and the L1/L2 caches are shared among programs using the same physical core. Therefore, a covert channel can be built among hyper-threads and threads sharing a logical core in a time-sliced setting. The LLC, cache coherence states, and DRAM are shared among different cores on the chip, and therefore, a covert channel can be built between different cores.

Some covert channels may use more than one component listed in Table~\ref{tbl:channel_share}. For example, in the cache hierarchy, there could be multiple levels of caches shared among the sender and the receiver. In Flush+Reload cache covert channel, the receiver can use the {\em clflush} instruction to flush a cache line from all the caches, and the sender may load the cache line into L1/L2 of that core or the shared LLC. If the sender and the receiver are in the same core, then the receiver will reload the data from L1. If the sender and the receiver are in different cores and only sharing the LLC, the receiver will reload the data from LLC.
Therefore, even with the same covert channel protocol, the location of the covert channel depends on the actual setting of the sender and the receiver.

As shown in Table~\ref{tbl:channel_share}, the channels in caches have relatively high bandwidth ($\sim$1MBits/s), which allows the attacker to launch efficient attacks. Covert channels in AVX and TLB are slower but enough for practical attacks.





\subsection{Disclosure Gadget}

\begin{figure}[t]
\includegraphics[width=3.2in]{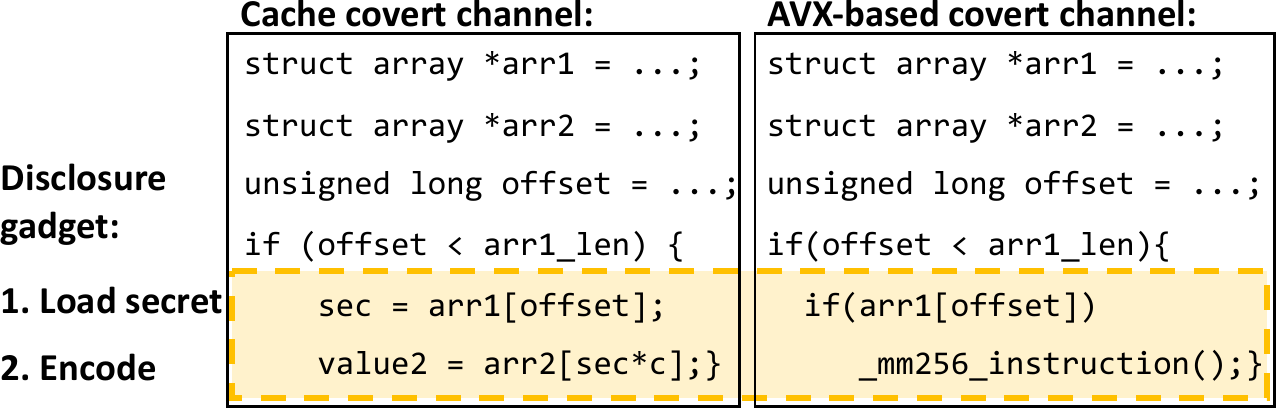}
\caption{\small Example disclosure gadgets for different covert channels.}
 \label{fig:disclosure_gadget}
\end{figure}

The covert channel is used in the disclosure gadget to transfer the secret to be accessible to the attacker architecturally. 
Disclosure gadget usually contains two steps: 1. load the secret into a register;  2. encode the secret into a covert channel.
As shown in Figure~\ref{fig:disclosure_gadget}, the disclosure gadget code depends on the covert channel used. For covert channels in the memory hierarchy (e.g., cache side channel), it will consist of memory access whose address depends on the secret value. For AVX-based covert channels, the disclosure gadget encodes the secret by using (or not using) AVX instruction.

%% file: transient_attacks.tex
\section{Existing Transient Execution Attacks}
\label{sec:transient_attacks}

The transient execution attacks contain two parts: triggering transient execution to obtain data 
that is otherwise not accessible (discussed in Section~\ref{sec:transient_exe}) 
and transferring the data via a covert channel (discussed in Section~\ref{sec:covert_channel}).
If the victim executes transiently, the victim will encode the secret into the channel, and the behavior cannot be analyzed from 
the software semantics without a hardware model of prediction.
If the attacker executes transiently, the micro-architecture propagates data that is not allowed to propagate at the ISA level (propagation is not visible at ISA level, but can be reconstructed through cover channels which observe the changes in micro-architecture).
To formally model and detect the behavior, a new micro-architectural model, including the transient behavior, should be used~\cite{mcilroy2019spectre,cryptoeprint:2019:310,guarnieri2018spectector,guarnieri2020hardware}.

\subsection{Existing Transient Execution Attacks Types}

\begin{table*}[t]
\centering
\caption{\small  Transient Execution Attacks Types.}
\begin{threeparttable}
\small
\begin{tabular}{  l | l l l l l l }
& \multicolumn{6}{c}{\bf Cause of Transient Execution}\\
 \parbox{0.9in}{\bf Covert Channel} &\parbox{0.4in}{\rotatebox{0}{PHT}} &  \parbox{0.4in}{\rotatebox{0}{BTB}}  &  \parbox{0.4in}{\rotatebox{0}{RSB}} &  \parbox{0.4in}{\rotatebox{0}{STL}}  & \parbox{0.4in}{\rotatebox{0}{LFB}} & \parbox{0.5in}{\rotatebox{0}{Exception}}  \\
\hline
 Execution Ports &$\Box$ &\cite{bhattacharyya2019smotherspectre} &$\Box$&$\Box$&$\Box$ &$\Box$\\
 L1 Cache Ports &$\Box$ &$\Box$&$\Box$&$\Box$ &$\Box$&$\Box$\\
 Memory Bus &$\Box$ &$\Box$&$\Box$&$\Box$ &$\Box$&$\Box$\\
 AVX2 unit &\cite{schwarz2018netspectre}& $\Box$ &$\Box$ &$\Box$&$\Box$&$\Box$ \\
 FP div unit  &\cite{fustos2020spectrerewind}& $\Box$ &$\Box$ &$\Box$&$\Box$&\cite{fustos2020spectrerewind} \\
TLB & $\Box$ &$\Box$&$\Box$&$\Box$ &$\Box$ &$\Box$\\
L1, L2 (tag, LRU) &\cite{Kocher2018spectre}& \cite{Kocher2018spectre,chen2018sgxpectre} & \cite{maisuradze2018ret2spec,koruyeh2018spectre}&\cite{v4,fallout} &\cite{ridl,Schwarz2019ZombieLoad} &\cite{Lipp2018meltdown,stecklina2018lazyfp,van2018foreshadow,weisse2018foreshadow,v3a,kiriansky2018speculative,van2020lvi}\\
LLC (tag, LRU) &$\Box$ &$\Box$&$\Box$&$\Box$ &$\Box$ &$\Box$\\
Cache Coherence  &\cite{trippel2018meltdownprime} &$\Box$ &$\Box$&$\Box$&$\Box$ &\cite{trippel2018meltdownprime}\\
Cache Directory &$\Box$ &$\Box$&$\Box$&$\Box$ &$\Box$ &$\Box$\\
DRAM row buffer &$\Box$ &$\Box$&$\Box$&$\Box$ &$\Box$ &$\Box$ \\
Other Channel &$\Box$ &$\Box$&$\Box$&$\Box$ &$\Box$ &$\Box$\\
\hline
\end{tabular}
\begin{tablenotes}
$\Box$ shows attacks that are possible but not demonstrated yet.
\end{tablenotes}
\end{threeparttable}
\label{tbl:attack_demo}
\end{table*}

To launch an attack, the attacker needs a way to cause transient execution of the victim or herself and a covert channel.
Table~\ref{tbl:attack_demo} shows the attacks that are demonstrated in the publications. For demonstrating different speculation primitives, researchers usually use the covert channel in caches (row L1, L2 in Table~\ref{tbl:attack_demo}). This is because the cache Flush+Reload covert channel is simple and efficient.
For demonstrating different covert channels used in transient execution attacks, researchers usually use PHT (Spectre V1).  This is because Spectre V1 is easy to demonstrate.
Note that every entry in the table can become an attack.
For mitigations, each entry of the table should be mitigated, either mitigate all the covert channels or prevent accessing the secret data in transient execution.

\subsection{Feasibility of Existing Attacks}

\subsubsection{Feasibility of the Transient Execution} 
As discussed in Section~\ref{sec:feasibility_scenario} and Section~\ref{sec:mistrain_setup}, Spectre attacks require the attacker to mis-train the prediction unit in the setup phase to let the victim execute gadget speculatively. To be able to mis-train, the attacker either needs to control part of the victim's execution to generate the desired history for prediction or needs to co-locate with the victim on the same core.
MDS attacks also require the attacker and the victim to share the same address speculation unit.
As shown in Table~\ref{tbl:pred_share}, the prediction unit is shared only within a physical core, for some unit, not even share between each hyper-thread. In practice, it is not trivial to co-locate on the same core.


\subsubsection{Feasibility of the Covert Channel}
As shown in Table~\ref{tbl:pred_ctrl_victim} and Section~\ref{sec:comp_CC}, in some scenarios, a covert channel across processes is required, and thus, the sharing of hardware is needed, which requires the co-location of threads. Furthermore, for a certain attack implementation, only one disclosure primitive is used, and the attack can be mitigated by blocking the covert channel.

\subsection{Attacks on Different Commercial Platforms}

\hl{Most of the existing studies focus on Intel processors, Table}~\ref{tbl:commercial}\hl{ lists the known attacks on processors by different venders, such as} AMD~\cite{AMD_attack,canella2018systematic}, Arm~\cite{canella2018systematic,ARM_attack}, RISC-V~\cite{gonzalez2019replicating}\hl{. As shown in the table, Spectre-type attacks using branch prediction are found on all the platforms, this is because branch speculation is fundamental in modern processors. Other transient execution depends on the micro-architecture implementation of speculation units, and show different results on different platforms.}

\begin{table*}[t]
\centering
\caption{\small \hl{Known Transient Execution Attacks on Different Platforms.}}
\begin{threeparttable}
\centering
\small
\begin{tabular}{ l l| c c c c c c c  }
 \multicolumn{2}{C{18em}|}{\bf Cause of Transient Execution}  &  {\bf Intel} &  {\bf AMD~\cite{AMD_attack,canella2018systematic}} & {\bf Arm~\cite{canella2018systematic,ARM_attack}} &  {\bf RISC-V\cite{gonzalez2019replicating}}  \\
\hline
\multirow{3}{6em}{{Control Flow}} 
& PHT (V1)& $\boxtimes$ & $\boxtimes$ &$\boxtimes$ & $\boxtimes$ \\
& BTB (V2)&$\boxtimes$ & $\boxtimes$ &$\boxtimes$ &$\boxtimes$ \\
& RSB (V5)&$\boxtimes$ & $\Box$\ &$\boxtimes$  & $\Box$\\
\hline
\multirow{2}{10em}{{Address Speculation}} 
& STL (V4,MDS) &$\boxtimes$ &$\Box$  &$\boxtimes$  &$\Box$\ \\
& LFB (MDS)&$\boxtimes$  &$\Box$  &$\Box$ &$\boxtimes$ \\
\hline
\multirow{6}{4em}{{Exception}} 
&  PF-US (V3) &$\boxtimes$&$\Box$ & $\boxtimes$ & $\Box$\\
& PF-P (L1TF) &$\boxtimes$ &$\Box$  &$\Box$ &$\Box$  \\
& PF-RW (V1.2) &$\boxtimes$ &$\Box$ &$\boxtimes$ &$\Box$  \\
& NM (LazyFP)  &$\boxtimes$ &$\Box$ &$\Box$ &$\Box$  \\
& GP (V3a) &$\boxtimes$&$\Box$ &$\boxtimes$  &$\Box$  \\
& Other & $\boxtimes$&$\boxtimes$ &$\boxtimes$  &$\Box$\\
\hline
\end{tabular}
\begin{tablenotes}
$\boxtimes$ indicates that an attack of the type on the platform; $\Box$ indicates that there is no known attack.
\end{tablenotes}
\end{threeparttable}
\label{tbl:commercial}
\end{table*}

%% file: mitigations.tex
\section{Mitigations of Spectre-type Attacks in Micro-architecture Design}
\label{sec:mitigations}

In this section, we focus on micro-architectural mitigations to attacks that occur when the victim executes transiently under wrong control flow prediction. 
\hl{As shown in Table}~\ref{tbl:commercial}\hl{, attacks that leveraging control flow prediction are more fundamental and affect all modern computer architectures.
Attacks that leveraging address speculation and exceptions are implementation-dependent, and we consider them as implementation bugs. They can be fixed, although the performance penalty is unknown now.}
We focus on possible future micro-architecture designs that are safe against control flow prediction. Thus, software mitigation schemes, such as~\cite{LoadHardening,oleksenko2018you,retpoline}, \hl{and software vulnerability detection schemes}~\cite{oleksenko2020specfuzz,wang2020kleespectre,wang2019oo7} are not discussed in detail.



\subsection{Mitigating Transient Execution}

The simplest mitigation is to stop any transient execution. However, it will come with a huge performance overhead, e.g., adding a fence after each branch to stop branch prediction causes 88\% performance loss~\cite{yan2018invisispec}.

\subsubsection{Mitigating the Trigger of Transient Execution}
To mitigate Spectre-type attacks, one solution is to limit the attackers' ability to mis-train the prediction units to prevent the disclosure gadget to be executed transiently (the first metric in Section~\ref{sec:transient_metric}). The prediction units (e.g., PHT, BTB, RSB, STL) should not be shared among different users. This can be achieved by static partition for concurrent users and flush the state during context switches.
For example, there are ISA extensions for controlling and stopping indirect branch predictions~\cite{AMD_mitigation,Intel_mitigation}. In~\cite{taram2019context}, a decode-level branch predictor isolation technique is proposed, where a special micro-op that clears the branch predictor states will be executed when the security domain switches. In~\cite{zhao2020lightweight}, it is proposed to use thread-private random number to encode the branch prediction table, to build isolation between threads in the branch predictor.
However, for both proposals, if the attacker can train the prediction unit by executing victim code with certain input (e.g., always provide valid input in Spectre V1), isolation is not enough. 

There is also mitigation in software to stop speculation by making the potential secret data depends on the result of the branch condition leveraging data dependency, e.g., masking the data with the branch condition~\cite{LoadHardening,oleksenko2018you}, because current processors do not speculate on data.
However, this solution requires to identify all control flow dependency and all disclosure gadgets, to figure out all possible control flow that could lead to the execution of the disclosure gadgets, and to patch each of them. 
It is a challenge to identify all (current and future) disclosure gadgets, because disclosure gadgets may vary due to the encoding to different covert channels, and formal methods that model the micro-architecture behavior are required~\cite{guarnieri2018spectector,guarnieri2020hardware}.


\subsubsection{Mitigating Transient Execution of Disclosure Gadget}


To mitigate leak of secret during the transient execution attacks, one way is to prevent the transient execution of the disclosure gadget, i.e., to stop loading of secrets in transient execution or stop propagating the secret to younger instructions in the disclosure gadget transiently.
For Meltdown-type and MDS-type attacks, it means to stop propagating secret data to the younger instructions. 
For Spectre-type attacks, however, the logic may not know which data is secret. 
To mitigate the attacks, secret data should be tagged with metadata as in secure architecture designs, which will be discussed in Section~\ref{sec:sec_arch}.

Another solution is that data cannot be propagated speculatively, and thus, cannot be send to covert channels speculatively, which can potentially prevent transient execution attacks with any covert channel.
In {\em Context-Sensitive Fencing}~\cite{taram2019context}, fences will be injected at decoder-level to stop speculative data propagation if there are potential Spectre attacks.
In {\em NDA}~\cite{weisse2019nda}, a set of propagation policies are designed for defending the attacks leveraging different types of transient executions (for example, transient execution due to branch prediction or all transient execution), showing the trade-off between security and performance.
Similarly, in {\em SpecShield}~\cite{barber2019specshield,barber2019isolating}, different propagation policies are designed and evaluated.
In {\em Conditional Speculation}~\cite{li2019conditional},  the authors propose a defense scheme targeting covert channels in the memory system and propose an architecture where data cannot be transiently propagated to instructions that lead to changes in memory system showing $13\%$ performance overhead. To reduce performance overhead of the defense, they further change the design to only target Flush+Reload cache side channels, resulting performance overhead of $7\%$.
Furthermore, in {\em STT}~\cite{yu2019speculative}, a dynamic information flow tracking based micro-architecture is proposed to stop the propagation of speculative data to covert channels but reduce the performance overhead by waking up instructions as early as possible.
\hl{Speculative data-oblivious (SDO) execution}~\cite{yu2020speculative}\hl{ is based on STT. To reduce performance overhead, SDO introduces new predictions that do not depend on operands (holding data potentially depending on speculative data). Specifically, speculative data-oblivious loads are designed to allow safe speculative load. }
The overhead to defend Spectre-like attacks is moderate, e.g., $7.7\%$ in  {\em Context-Sensitive Fencing}~\cite{taram2019context}, $21\%$ reported in {\em SpecShield}~\cite{barber2019specshield},  $20\sim51\%$ ($113\%$ for defending all transient execution attacks) reported in {\em NDA}~\cite{weisse2019nda}, and $8.5\%$ for branch speculation ($14.5\%$ for all transient execution) in {\em STT}~\cite{yu2019speculative}, $4.19\%$ for branch speculation ($10.05\%$ for all transient execution) in {\em STT+SDO}~\cite{yu2020speculative}.

\begin{table*}[t]
\centering
\caption{\small Comparison of Different Mitigation Schemes in Micro-architecture.}
\begin{threeparttable}
\small
\begin{tabular}{   | p{2.2in} | p{3in} |  }
\hline
\textbf{Mitigation Schemes} & \textbf{Performance Overhead} \\
\hline
Fence after each branch & $88\%$~\cite{yan2018invisispec} \\
\hline
Stop propagating all data & $30$--$55\%$~\cite{barber2019isolating}; $21\%$~\cite{barber2019specshield}; $20$--$51\%$~\cite{weisse2019nda}; $8.5\%$~\cite{yu2019speculative}; $4.19\%$~\cite{yu2020speculative}  \\
Stop propagating all data to cache changes& $7.7\%$~\cite{taram2019context}, $13\%$~\cite{li2019conditional}\\
Stop propagating all data to Flush+Reload &$7\%$~\cite{li2019conditional}\\
\hline
Stop propagating all tagged secret data& $71\%$ for security-critical applications, $<1\%$~for real-world workloads~\cite{schwarz2019context,fustos2019spectreguard}\\
\hline
Partitioned cache & $1$--$15\%$~\cite{kiriansky2018dawg} \\
\hline
Stop (Undo) speculative change in caches &7.6\%~\cite{yan2018invisispec}; 11\%~\cite{sakalis2019efficient}; 4\%~\cite{ainsworth2019muontrap};  5.1\%~\cite{saileshwar2019cleanupspec}; 8.3\%~\cite{wu2020reversispec}\\
\hline
\end{tabular}
\begin{tablenotes}
\end{tablenotes}
\end{threeparttable}
\label{tbl:mitigation}
\end{table*}

There should be a large enough speculative window to let the disclosure gadget execute transiently for the attack to happen. The micro-architecture may be able to limit the speculation window size to prevent the encoding to the covert channel (the fourth metric in Section~\ref{sec:transient_metric}). However, the disclosure gadget can be very small that only contains two loads from L1~\cite{Xiong2019Leaking}, which is only about 20 cycles in total. Detecting a malicious windowing gadget accurately can be~challenging.

\subsubsection{Mitigations in Secure Architectures}
\label{sec:sec_arch}
Secure architectures are designed to protect the confidentiality (or integrity) of certain data or code.
Thus, secure architectures usually come with ISA extensions to identify the data or code to be protected, e.g., secret data region, and micro-architecture designs to isolate the data and code to be protected~\cite{costan2016intel,suh2014aegis,lie2000architectural}.

With knowledge about the data to be protected, hardware can further stop propagating secret data during speculation.
The hardware can identify data that is depended on the secret with taint checking, as proposed in~\cite{Kocher2018spectre,schwarz2019context,fustos2019spectreguard,taram2019context}, and forbid tainted data to have micro-architectural side effects, or flush
all the states on exit from the protected domain, to defend against persistent covert channels, and disable SMT to defend volatile covert channels.
The overhead of such mitigation depends on the size of secret data to be protected. For example, as reported in {\em ConTExT}~\cite{schwarz2019context}, the overhead is $71.14\%$ for OpenSSL RSA encryption and less than $1\%$ for real-world workloads.
Similar overhead is reported in {\em SpectreGuard}~\cite{fustos2019spectreguard}.
Intel also proposed a new memory type, named speculative-access protected memory (SAPM)~\cite{SAPM}. Any access to SAPM region will cause instruction-level serialization and speculative execution beyond the SAPM-accessing instruction will be stopped until the retirement of that instruction.

\subsection{Mitigating Covert Channels}

To limit the covert channels, one way is to isolate all the hardware across the sender and receiver of the channel, so the change cannot be observable to the receiver. However, this is not always possible, e.g., in some attacks, the attacker is both the sender and the receiver of the channel.

Another mitigation is to eliminate the sender of the covert channel in transient execution.
For volatile covert channels, the mitigation is challenging.
For permanent covert channels, there should not be speculative change to any micro-architectural states or any micro-architectural state changes should be rolled back when the pipeline is squashed. 
Covert channels in memory systems, such as caches and TLBs, are most commonly used. Hence, most of the existing mitigations focus on cache and TLB side channels.

{\em InvisiSpec}~\cite{yan2018invisispec} proposed the concept of ``visibility point" of a load, which indicates the time when a load is safe to cause micro-architecture state changes that are visible to attackers. Before the visibility point, a load may be squashed, and should not cause any micro-architecture state changes visible to the attackers. To reduce performance overhead, a ``speculative buffer" is used to temporarily cache the load, without modifications in the local cache. After the ``visibility point", the data will be fetched into the cache. For cache coherency, a new coherency policy is designed such that the data will be validated when stale data is potentially fetched. The gem5~\cite{binkert2011gem5} simulation results show a 7.6\% performance loss for SPEC 2006 benchmark~\cite{henning2006spec}.
Similarly, SafeSpec~\cite{khasawneh2018safespec} proposed to add ``shadow buffers" to caches and TLBs, so that transient changes in the caches and TLBs does not happen. 

In {\em Muontrap}~\cite{ainsworth2019muontrap}\hl{, ``filter cache" (L0 cache) is added to each physical thread to hold speculative data. 
The proposed filter cache only holds data that is in Shared state, so it will not change the timing of accessing other caches. If the shared state in L0 is not possible without causing the cache line in another cache to change state form Modified or Exclusive state, the access will be delayed until it is at the head of ROB.
The cache line will be written through to L1 when the corresponding instruction commits.
Different from the buffers in InvisiSpec}~\cite{yan2018invisispec} and SafeSpec~\cite{khasawneh2018safespec}\hl{, the filter cache is a real cache that is cleared upon a context switch, syscall, or when the execution change security boundaries (e.g., explicit flush when exiting sandbox) to ensure isolation between security boundaries. Muontrap results in a 4\% slowdown for SPEC 2006.}

{\em CleanupSpec}~\cite{saileshwar2019cleanupspec} proposed to use a combination of undoing the speculative changes and secure cache designs. When mis-speculation is detected and the pipeline is squashed, the changes to the L1 cache are rolled back. For tracking the speculative changes in caches, 1Kbyte storage overhead is introduced.
To prevent the cross-core or multi-thread covert channel, partitioned L1 with random replacement policy and randomized L2/LLC are used.
Because only a small portion of transient executions results in mis-speculations, the method shows an average slowdown of 5.1\%.

{\em ReversiSpec}~\cite{wu2020reversispec} \hl{proposed a comprehensive cache coherence protocol considering speculative cache accesses. The cache coherence protocol proposed an interface including three operations: 1) speculative load, 2) merge when a speculative load is safe, 3) purge when a speculative load is squashed. Compared to InvisiSpec}~\cite{yan2018invisispec}\hl{, the speculative buffer only stores data when the data is not in the cache, and thus,
less data movement will occur when a load is safe (merge). Compared to CleanupSpec}~\cite{saileshwar2019cleanupspec}\hl{,  purge is fast as not all the changes have propagated in to cache. The performance overhead is 8.3\%.}

Moreover, accessing speculative loads that hit in L1 cache will not cause side effects (except LRU state updates) in the memory system. Therefore, only allowing speculative L1 hits can mitigate transient execution attacks using covert channels (other than LRU) in the memory system. 
In {\em Selective Delay}~\cite{sakalis2019efficient}, to improve performance, for a speculative load that miss in L1, value prediction is used. The load will fetch from deeper layers in the memory hierarchy until the load is not speculative. In their solution, 11\% performance overhead is shown.

Meanwhile, many secure cache architectures are proposed to use randomization to mitigate the cache covert channels in general (not only the transient execution attacks).
For example, {\em Random Fill cache}~\cite{liu2014random} decouples the load and the data that is filled into the cache, and thus, the cache state will no longer reflect the sender's memory access pattern.
{\em Random Permutation (RP) cache}~\cite{wang2007new}, {\em Newcache cache}~\cite{wang2008novel,liu2016newcache}, {\em CEASER cache}~\cite{qureshi2018ceaser}, and {\em ScatterCache}~\cite{werner2019scattercache} randomize memory-to-cache-set mapping to mitigate contention-based occupancy-based covert channels in the cache.
{\em Non Deterministic cache}~\cite{keramidas2008non} randomizes cache access delay and de-couple the
relation between cache block access and cache access timing.
Secure TLBs~\cite{deng2019secure} are also proposed to mitigate covert channels in TLBs.
But again, all the possible covert channels need to be mitigated to fully mitigate transient execution attacks.
Further, {\em Cyclone}~\cite{harris2019cyclone} proposed a micro-architecture to detect cache information leaks across security domains.

Another mitigation is to degrade the quality of the channel or even make the channel unusable for a practical attack.
For example, many timing covert channels require the receiver to have a fine-grained clock to observe the channel (the second metric in Section~\ref{sec:CC_metric}). Limiting the receiver's observation will reduce the bandwidth or even mitigate the covert channel~\cite{Webkit,schwarz2018javascript}.  Noise can also be added to the channel to reduce the bandwidth (the third metric in Section~\ref{sec:CC_metric}).

However, the above mitigations only cover covert channels in memory systems.
To mitigate other covert channels, there are the following challenges:
1. Identify all possible covert channels in micro-architecture, including future covert channels.
Formal methods are required in this process. For example, information flow tracking, such as methods in~\cite{zhang2015hardware,zhang2012language,deng2019secchisel}, can be used to analyze the hardware components, where the data of transient execution could flow to. Then, analyze if each of the components could result in a permanent or transient covert channel.
2. Mitigate each of the possible covert channels.

\subsubsection{Mitigations in Secure Architectures}
\label{sec:sec_arch_covert_channel}
With clearly defined security domain, isolation can be designed to mitigate not only transient covert channels and also conventional covert channels.
For example, to defend cache covert channels, a number of partitioned caches to different security domains are proposed, either statically~\cite{lee2005architecture,he2017secure,zhang2015hardware,zhang2012language,kiriansky2018dawg,yan2017secure,costan2016sanctum,bourgeat2018mi6,liu2016catalyst,wang2007new} or dynamically~\cite{wang2016secdcp,domnitser2012non}. With partition, shared resource no longer exists between the sender and the receiver, and the receiver cannot observe secret dependent behavior to decode the secret.

The above proposal assumes the hardware is isolated for each security domain. However, there is also a scenario where software outside the security domain may use the same hardware after a context switch.
In {\em Mi6} processor~\cite{bourgeat2018mi6}, caches and ports partitioning are used to isolate software on different cores. Further, when there is a context switch, a security monitor flushes the architecture and micro-architecture states, which holds the information of in-flight speculation from the previously executing program. To protect the security monitor, speculation is not used in the execution of the security monitor. In OPTIMUS~\cite{omar2020optimus}\hl{, a dynamic partitioning scheme in the granularity of core is proposed to achieve both security and high performance.}

%% file: conclusion.tex
\section{Conclusion}
\label{sec:conclusion}
This paper provide a survey of the transient execution attacks.
This paper first defines the transient execution attacks and the three phases of the attacks. 
It then categorizes possible causes of transient executions. 
The security boundaries that are broken in the transient executions are discussed.
It also analyzes the causes of transient execution by proposing a set of metrics and using the metrics to compare the feasibility. 
Furthermore, the covert channels that can be used in the attacks are categorized and compared with a new set of metrics.
Combining the transient execution and the covert channels, different types of attacks are compared. In the end, possible mitigation schemes in micro-architecture designs are discussed and compared.